\documentclass[twocolumn,showpacs,aps,prc,superscriptaddress]{revtex4}

\usepackage[dvips]{graphicx}
\usepackage{color}
\usepackage{latexsym}

\newcommand{\epg}{e p \to e p \gamma}
\newcommand{\xvertex}{x_{v}}
\newcommand{\yvertex}{y_{v}}
\newcommand{\zvertex}{z_{v}}
\newcommand{\twoarmx}{x_{2arms}}
\newcommand{\ddd}{x_{dif}}
\newcommand{\plltt}{P_{LL}-P_{TT}/\epsilon}
\newcommand{\plt}{P_{LT}}
\newcommand{\pll}{P_{LL}}
\newcommand{\ptt}{P_{TT}}
\newcommand{\ale}{\alpha_E}
\newcommand{\bem}{\beta_M}
\newcommand{\lama}{\Lambda_{\alpha}}
\newcommand{\lamb}{\Lambda_{\beta}}
\newcommand{\hrse}{HRS-E }
\newcommand{\hrsh}{HRS-H }
\newcommand{\hrsenospace}{HRS-E}
\newcommand{\hrshnospace}{HRS-H}
\newcommand{\nucleonmass}{m_p}
\newcommand{\pionmass}{m_{\pi^0}}
\newcommand{\ycollih}{y_{collim}^{Hadron}}

\newcommand{\CM}{CM }
\newcommand{\CMnospace}{CM}

\begin{document}

%
%
%
%

\title{
Virtual Compton Scattering and the Generalized Polarizabilities of the Proton at $Q^2$=0.92 and 1.76 GeV$^2$ 
}

%
%
%
%

\author{H.~Fonvieille \footnote{corresponding author, helene@clermont.in2p3.fr}}
\affiliation{Clermont-Universit\'e, UBP, CNRS-IN2P3, LPC, BP 10448, F-63000 Clermont-Ferrand, France}
\author{G.~Laveissi\`{e}re}
\affiliation{Clermont-Universit\'e, UBP, CNRS-IN2P3, LPC, BP 10448, F-63000 Clermont-Ferrand, France}
\author{N.~Degrande}
\affiliation{Department of Physics and Astronomy, Ghent University, B-9000 Ghent, Belgium}
\author{S.~Jaminion}
\affiliation{Clermont-Universit\'e, UBP, CNRS-IN2P3, LPC, BP 10448, F-63000 Clermont-Ferrand, France}
\author{C.~Jutier}
\affiliation{Clermont-Universit\'e, UBP, CNRS-IN2P3, LPC, BP 10448, F-63000 Clermont-Ferrand, France}
\affiliation{Old Dominion University, Norfolk, VA 23529}
\author{L.~Todor}
\affiliation{Old Dominion University, Norfolk, VA 23529}
\author{R.~Di Salvo}
\affiliation{Clermont-Universit\'e, UBP, CNRS-IN2P3, LPC, BP 10448, F-63000 Clermont-Ferrand, France}
\author{L.~Van Hoorebeke}
\affiliation{Department of Physics and Astronomy, Ghent University, B-9000 Ghent, Belgium}
\author{L.C.~Alexa}
\affiliation{University of Regina, Regina, SK S4S OA2, Canada}
\author{B.D.~Anderson}
\affiliation{Kent State University, Kent OH 44242}
% ...
\collaboration{The Jefferson Lab Hall A Collaboration}
\noaffiliation

%
%
%
%
%
%

\begin{abstract}
Virtual Compton Scattering (VCS) on the proton has been studied at Jefferson Lab using the exclusive photon electroproduction reaction $ep \to ep \gamma$. This paper gives a detailed account of the analysis which has led to the determination of the structure functions $\plltt$ and $\plt$, and the electric and magnetic generalized polarizabilities (GPs) $\ale (Q^2)$ and $\bem (Q^2)$ at values of the four-momentum transfer squared Q$^2$= 0.92 and 1.76 GeV$^2$. These data, together with the results of VCS experiments at lower momenta, help building a coherent picture of the electric and magnetic GPs of the proton over the full measured $Q^2$-range, and point to their non-trivial behavior.
\end{abstract}

\maketitle


\section{Introduction} \label{sec-intro}

The nucleon is a composite object, and understanding its structure is the subject of intensive efforts. Its electromagnetic structure is cleanly probed by real and virtual photons. Real Compton scattering (RCS) at low energy gives access to the nucleon polarizabilities, which describe how the charge, magnetization and spin densities in the nucleon  are deformed when the particle is subjected to an external quasi-static electromagnetic field. 

Virtual Compton scattering  (VCS)  \ $\gamma^* N \to \gamma N$ \  gives access to the generalized polarizabilities (GPs). Being dependent on the photon virtuality $Q^2$, these observables parametrize the {\it local} polarizability response of the system, i.e. they give information on the density of polarization inside the nucleon. Experimental information on the GPs is obtained through the reaction of exclusive photon electroproduction. Several dedicated experiments on the proton:
$$  e \, p \ \to \ e \, p \, \gamma \ \ \ \ \ \ \ \ \ \ \ \ (1)  $$
have been performed at various $Q^2$ and in the low-energy regime. This includes the near-threshold  region, where the center-of-mass energy $W$ of the $(\gamma p)$ system is below the one-pion threshold ($W<( \nucleonmass + \pionmass )$, where $\nucleonmass$ and $\pionmass$ are the proton and pion masses), and up to the $\Delta(1232)$ resonance region. Process (1) has been studied experimentally  at MAMI~\cite{Roche:2000ng,Bensafa:2006wr,Janssens:2008qe,Sparveris:2008jx}, the Thomas Jefferson National Accelerator Facility (JLab)~\cite{Laveissiere:2004nf,Laveissiere:2008zn} and MIT-Bates~\cite{Bourgeois:2006js,Bourgeois:2011te}.


The results of the near-threshold VCS data analysis of the JLab VCS experiment E93-050, i.e. the structure functions $\plltt$ and $\plt$, and the electric and magnetic GPs $\ale$ and $\bem$ at $Q^2$= 0.92 and 1.76 GeV$^2$, have been published elsewhere \cite{Laveissiere:2004nf}. However, analysis details and cross section data were not given. This is the aim of the present paper, which is organized as follows. After recalling briefly the theoretical concepts  in section~\ref{sec-theo}, the experimental setup is described in section~\ref{sec-experiment}. Section~\ref{sect-data-ana} reports about data analysis, including event reconstruction, acceptance calculation and cross section determination. Section~\ref{sec-resul} presents the measured cross section, the physics results deduced from the various analyses, and a discussion. A short conclusion ends the paper in section \ref{sec-concl}.

\section{Theoretical concepts and tools for experiments}\label{sec-theo}

This section summarizes the theoretical concepts underlying the measurements of VCS at low energy: the GPs, the structure functions, and the principles of measurement. For details, we refer to review papers: \cite{Guichon:1998xv,Drechsel:2002ar,Pasquini:2011zz} (theory) and \cite{d'Hose:2000xr,HydeWright:2004gh,Fonvieille:2004rb,d'Hose:2006xz,Downie:2011mm} (experiments).

\subsection{Generalized polarizabilities}\label{subsec-gp}

Polarizabilities are fundamental characteristics of any composite system, from hadrons to atoms and  molecules. They describe how the system responds to an external quasi-static electromagnetic field. Real Compton Scattering (RCS) yields for the static polarizabilities of the proton  \cite{OlmosdeLeon:2001zn}:
\begin{tabular}{llll}
$\ale$ (electric) & = & $( 12.1$ & $ \pm 0.3_{stat} \mp 0.4_{syst} ) \cdot 10^{-4} fm^3 $ \\
$\bem$ (magnetic)  & = & $( 1.6$  & $ \pm 0.4_{stat} \pm 0.4_{syst} ) \cdot 10^{-4} fm^3 .  $ \\
\end{tabular} \newline
These values are much smaller than the particle's volume and indicate that the proton is a very stiff object, due to the strong binding between its constituents.

The formalism of VCS on the nucleon was early explored in~\cite{BergLindner:1961} and the concept of generalized polarizabilities was first introduced in~\cite{DrechselArenhovel:1974} for nuclei. The nucleon case was established within a Low Energy Theorem (LET), first applied to VCS by P. Guichon {\it et al.} in~\cite{Guichon:1995pu}. This development paved the way to new experimental investigations: it became possible to explore the spatial distribution of the nucleon's polarizability response, which is in essence the physical meaning of the GPs (see e.g. \cite{Scherer:2004fd,Gorchtein:2009qq}).

\begin{figure}[htb]
\includegraphics[width=8.3cm]{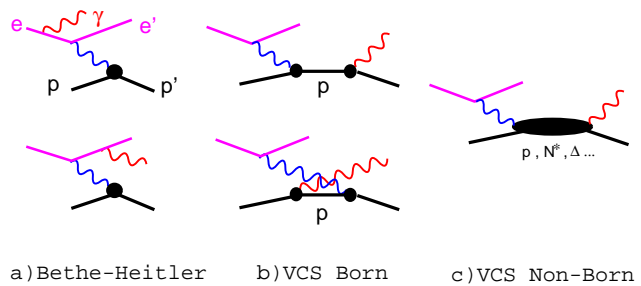}  
\caption{(Color online) Feynman graphs of photon electroproduction.}
\label{fig-bh+b+nb}
\end{figure}

Photon electroproduction accesses VCS via the amplitude decomposition shown in Fig.~\ref{fig-bh+b+nb}:    $T^{e p \gamma} = T^{BH} + T^{VCSBorn} + T^{VCSNonBorn}$, where BH stands for the Bethe-Heitler process. Formally the GPs  are obtained from the multipole decomposition of the Non-Born amplitude $T^{VCSNonBorn}$ taken in the ``static field'' limit $q'_{c.m.} \to 0$, where $q'_{c.m.}$ is the momentum of the final real photon in the $\gamma p$ center-of-mass (noted \CM hereafter). The GPs are functions of $q_{c.m.}$, the  momentum of the virtual photon in the \CMnospace, or equivalently the photon virtuality  $Q^2$ (see Appendix \ref{app-1} for more details). After the work of Drechsel {\it et al.} \cite{Drechsel:1998xv,Drechsel:1998zm}, six independent GPs remain at lowest order. Their standard choice is given in  Table \ref{tab-dipole-gps}, where they are indexed by the EM transitions involved in the Compton process. Since this paper mainly focuses on the electric and magnetic GPs, i.e. the two scalar ones (or spin-independent, or non spin-flip, S=0; see Table \ref{tab-dipole-gps}), we recall their definition:
\begin{eqnarray*}
\begin{array}{lll}
\ale (Q^2) & = & - P^{(L1,L1)0} (Q^2) \cdot ( {e^2 \over 4 \pi} \sqrt{3 \over 2} )  \ , \\
\bem  (Q^2) & = & - P^{(M1,M1)0} (Q^2) \cdot ( {e^2 \over 4 \pi} \sqrt{3 \over 8} ) \ .  \\
\end{array}
\end{eqnarray*}
These GPs coincide in the limit $Q^2 \to 0$ with the usual static RCS polarizabilities $\ale$ and $\bem$ introduced above.

\begin{table}[htbp]
\begin{center}
\caption{The standard choice for the nucleon GPs. In the notation of the first column,  $\rho (\rho')$ refers to the magnetic (1) or longitudinal (0) nature of the initial (final) photon, $L (L')$ represents the angular momentum of the initial (final) photon, and  $S$ differentiates between the spin-flip $(S=1)$ and non spin-flip $(S=0)$ character of the transition at the nucleon side. The multipole notation in the second column uses the magnetic (M) and longitudinal (L) multipoles. The six listed GPs correspond to the lowest possible order in $q'_{c.m.}$, i.e. a  dipole final transition $(L'=1)$. The third column gives the correspondence in the RCS limit ($Q^2 \to 0$ or $q_{c.m.} \to 0$).
}
\label{tab-dipole-gps}
\begin{tabular}{ccc}
\hline
\hline
\  $P^{(\rho ' L', \rho L ) S } (q_{c.m.})$ \ & \ $P^{(f,i)S} (q_{c.m.}) \ $ 
& RCS limit  \\
\hline
 $P^{(01,01)0}$ & $P^{(L1,L1)0}$ & 
$ - {4 \pi \over e^2} \sqrt{2 \over 3} \ $ \boldmath $\alpha_E $   \\
 $P^{(11,11)0}$ & $P^{(M1,M1)0}$ & 
$ - {4 \pi \over e^2} \sqrt{8 \over 3} \ $ \boldmath $ \beta_M $  \\
 $P^{(01,01)1}$ & $P^{(L1,L1)1}$ & 0 \\
 $P^{(11,11)1}$ & $P^{(M1,M1)1}$ & 0  \\
 $P^{(01,12)1}$ & $P^{(L1,M2)1}$ & 
$ - {4 \pi \over e^2} {\sqrt{2} \over 3} \ \gamma_3$  \\
 $P^{(11,02)1}$ & $P^{(M1,L2)1}$ & 
$ - {4 \pi \over e^2} { 2 \sqrt{2} \over 3 \sqrt{3}}  (\gamma_2 + \gamma_4)$ \\
\hline
\hline
\end{tabular}
\end{center}
\end{table}

\subsection{Theoretical models and predictions}\label{subsec-theo-models}

There are a number of theoretical models  which describe and calculate the GPs of the nucleon. They include: 
heavy baryon chiral perturbation theory (HBChPT)~\cite{Hemmert:1997at,Hemmert:1996gr,Hemmert:1999pz}, 
non-relativistic quark constituent models~\cite{Guichon:1995pu,Liu:1996xd,Pasquini:2000ue,Pasquini:2001pj}, 
dispersion relations~\cite{Pasquini:2000pk,Pasquini:2001yy,Drechsel:2002ar}, 
linear-$\sigma$ model~\cite{Metz:1996fn,Metz:1997fr}, 
effective Lagrangian model~\cite{Vanderhaeghen:1996iz}, 
Skyrme model~\cite{Kim:1997hq}, 
the covariant framework of ref.\cite{L'vov:2001fz}, 
or more recent works regarding 
GPs redefinition~\cite{Gorchtein:2009wz}, 
manifestly Lorentz-invariant baryon ChPT\cite{Dalibor:2008th}, 
or light-front interpretation of GPs~\cite{Gorchtein:2009qq}.

One of the main physical interests of GPs is that they can be sensitive in a specific way to the various physical degrees of freedom, e.g. the nucleon core and the meson cloud. Thus their knowledge can bring novel information about nucleon structure. The electric GP is usually predicted to have a smooth fall-off with $Q^2$. The magnetic GP has two contributions, of paramagnetic and diamagnetic origin; they nearly cancel, making the total magnitude small. As will be shown in section \ref{subsec-discuss-gp}, the available data more or less confirm these trends. A synthesis of diverse GP predictions for the proton is presented in \cite{Pasquini:2001pj}.

\subsection{The Low Energy Theorem and the structure functions}\label{subsec-lex-theo}

\begin{figure}[htb]
\includegraphics[width=8.3cm]{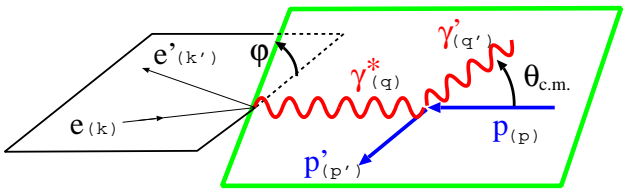}  
\caption{(Color online) $(ep \to ep \gamma)$ kinematics; four-momentum vectors notation and Compton angles $(\theta_{c.m.},\varphi)$ in the $\gamma p$ center-of-mass.
}
\label{fig-epgamma-kinem}
\end{figure}

The LET established in \cite{Guichon:1995pu} is a major tool for analyzing VCS experiments. The LET describes the photon electroproduction cross section below the pion threshold in terms of GPs. The (unpolarized) $\epg$ cross section at small $q'_{c.m.}$ is written as:
\begin{eqnarray}
 d^5 \sigma =    
 d^5 \sigma ^{BH+Born}  \ + \ q'_{c.m.} \cdot \phi \cdot \Psi_0 \ + \ {\cal O}(q'^2_{c.m.}) \ .  
\label{eq-let1} 
\end{eqnarray} 
The notation   $d^5 \sigma$ stands for $d^5 \sigma / dk'_{elab} d \Omega'_{e lab} d \Omega_{C.M.}$ where $k'_{elab}$ is the scattered electron momentum, $d \Omega'_{e lab}$ its solid angle and  $d \Omega_{\gamma c.m.}$  the solid angle of the outgoing photon (or proton) in the \CMnospace; $\phi$ is a phase-space factor. The (BH+Born) cross section is known and calculable, with the proton electromagnetic form factors $G_E^p$ and $G_M^p$ as inputs. The $\Psi_0$ term represents the leading polarizability effect. It is given by:
\begin{eqnarray}
\Psi_0 = v_1 \cdot 
(P_{LL} - {\displaystyle 1 \over \displaystyle \epsilon} P_{TT}) 
\ + \ v_2 \cdot  P_{LT} \ ,  
\label{eq-let2} 
\end{eqnarray} 
where $P_{LL}, P_{TT}$ and $\plt$ are three unknown structure functions, containing the GPs.  $\epsilon$ is the usual virtual photon polarization parameter and $v_1, v_2$ are kinematical coefficients depending on $(q_{c.m.},\epsilon,\theta_{c.m.},\varphi)$ (see \cite{Guichon:1998xv} for their full definition). The incoming photon is chosen to point in the $z$-direction. The Compton angles are the polar angle $\theta_{c.m.}$ of the outgoing photon in the \CM and the azimuthal angle $\varphi$ between the leptonic and hadronic planes; see  Fig.~\ref{fig-epgamma-kinem}.

 The expressions of the structure functions  useful to the present analysis are summarized here:
\begin{eqnarray}
\begin{array}{lll}
P_{LL}  & = &   
{ 4 \nucleonmass  \over \alpha_{em} } \cdot 
G_E^p(Q^2)\cdot  \alpha_E(Q^2) \ ,  \\
P_{TT}  & = & [P_{TTspin}]  \ ,    \\
P_{LT}  & = & - { 2 \nucleonmass  \over \alpha_{em} } 
 \sqrt{ {q_{c.m.}^2 \over Q^2} } \cdot 
G_E^p(Q^2) \cdot \beta_M(Q^2) +  [P_{LTspin}]  \ , \label{eq-sf12} 
\end{array}
\end{eqnarray} 
where $\alpha_{em}$ is the fine structure constant. The terms in square brackets are the spin part of the structure functions (i.e. containing only spin GPs) and the other terms are the scalar parts. The important point is that the electric and magnetic GPs enter only in $\pll$ and in the  scalar part of $\plt$, respectively.

%
%
%

In an unpolarized experiment at fixed $Q^2$ and fixed $\epsilon$, such as ours, only two observables can be determined using the LET:  $\plltt$ and $\plt$, i.e. only two specific combinations of GPs. 
%
%
%
To further disentangle the GPs, one can in principle make an $\epsilon$-separation of $\pll$ and $\ptt$ (although difficult to achieve),  
and in order to extract all individual GPs one has to resort to double polarization~\cite{Vanderhaeghen:1997bx}. Here we perform a LET, or LEX (for Low-energy EXpansion) analysis in the following way: the  two  structure functions $\plltt$ and $\plt$ are extracted  by a linear fit of the difference $d^5 \sigma^{exp} - d^5 \sigma^{BH+Born}$, based on eqs.(\ref{eq-let1}) and (\ref{eq-let2}), and assuming the validity of the truncation of the expansion to ${\cal O}(q'^2_{c.m.})$. Then, to further isolate the scalar part in these structure functions, i.e. to access $\ale (Q^2)$ and $\bem (Q^2)$, a model input is required, since the spin part is not known experimentally.

\subsection{The Dispersion Relations model} \label{subsec-dr-theo}

The Dispersion Relation (DR) approach is the second tool for analyzing VCS experiments. It is of particular importance in our case, so we briefly review its properties in this section. The DR formalism was developed by B.Pasquini {\it et al.} \cite{Pasquini:2001yy,Drechsel:2002ar} for RCS and VCS.  Contrary to the LET which is limited to the energy region below the pion threshold, the DR formalism is also valid in the energy region up to the $\Delta(1232)$ resonance -an advantage fully exploited in our experiment.


The Compton tensor is parametrized through twelve invariant amplitudes $F_{i} (i=1, \cdots, 12)$. The GPs are expressed in terms of the non-Born part  $F_i^{NB}$ of these amplitudes at the point $t=-Q^2, \nu=(s-u)/4 \nucleonmass  =0$, where $s,t,u$ are the Mandelstam variables of the Compton scattering. The  $F_i^{NB}$ amplitudes, except for two of them, fulfill unsubtracted dispersion relations. When working in the energy region up to the $\Delta(1232)$, these $s$-channel integrals are considered to be saturated by the $\pi N$ intermediate states. In practice, the calculation uses the MAID  pion photo- and electroproduction multipoles \cite{Drechsel:1998hk}, which include both resonant and non-resonant production mechanisms.

%
%
%

The amplitudes $F_1$ and $F_5$ have an unconstrained part, corresponding to asymptotic contributions and dispersive contributions beyond $\pi N$. For $F_5$ this part is dominated by the $t$-channel $\pi^0$ exchange; with this input, all four spin GPs are fixed in the model. For $F_1$, a main feature is that in the limit $(t=-Q^2, \nu=0)$ its non-Born part is proportional to the magnetic GP. The unconstrained part of $F _{1}^{NB}$ is estimated by an energy-independent function noted $\Delta \beta$, and phenomenologically associated with the  $t$-channel $\sigma$-meson exchange. This leads to the expression:
\begin{eqnarray}
 \bem (Q^2)  = \beta^{\pi N}(Q^2) +  \Delta \beta(Q^2) \ , 
\label{eq-dr-beta-00}
\end{eqnarray}
where $\beta^{\pi N}$ is the dispersive contribution calculated using MAID multipoles. The $\Delta \beta$ term is parametrized by a dipole form:
\begin{eqnarray}
 \Delta \beta   =  { \displaystyle [ \beta^{exp}   -  
\beta^{\pi N} ]_{Q^2=0} 
\over
\displaystyle ( 1 + Q^2/ \lamb^2 )^2 } \ . 
\label{eq-dr-beta-0}
\end{eqnarray}
An unconstrained part is considered also for a third amplitude, $F_2$. Since in the limit  $(t=-Q^2, \nu=0)$ the non-Born part of $F_2$ is proportional to the sum $( \ale + \bem )$, one finally ends with a decomposition similar to  eq.(\ref{eq-dr-beta-0}) for the electric GP itself:
\begin{eqnarray}
\begin{array}{llll}
\ &   \ale (Q^2) & =  &  \alpha^{\pi N}(Q^2) +  \Delta \alpha(Q^2) \ , \\ 
 \mbox{with} & \Delta \alpha & = & 
{ \displaystyle [ \alpha^{exp} - \alpha^{\pi N} ]_{Q^2=0} 
\over
\displaystyle ( 1 + Q^2/ \lama^2 )^2 } \ .
\label{eq-dr-alpha-0}
\end{array}
\end{eqnarray}


The implication for experiments is that in the DR model the two scalar GPs are not fixed. They depend on the free parameters $\lama$ and $\lamb$ (dipole masses),  which can be fitted from data. It must be noted that the choice of a dipole form in eqs.(\ref{eq-dr-beta-0}) and (\ref{eq-dr-alpha-0}) is arbitrary:  $\lama$ and  $\lamb$ only play the role of intermediate quantities in order to extract VCS observables, with minimal model-dependence. These parameters are not imposed to be constant with $Q^2$.  Our experimental DR analysis consists in adjusting $\lama$ and $\lamb$ by a fit to the measured cross section, separately in our two $Q^2$-ranges.  Then, in each $Q^2$-range the model is entirely constrained; it provides all VCS observables, at a given value of $Q^2$ representative of the range: the scalar GPs as well as the structure functions, in particular $\plltt$ and $\plt$.


\section{The experiment} \label{sec-experiment}

The photon electroproduction cross section is small and requires a high-performance equipment to be measured with accuracy. To ensure the exclusivity of the reaction, one must detect at least two of the three particles in the final state. The chosen technique is to perform electron scattering at high luminosity on a dense proton target, and detect in coincidence the two outgoing charged particles in magnetic spectrometers of high resolution and small solid angle. These devices ensure a clean detection and a good identification of process (1). Section~\ref{sec-experiment} describes how the experiment was designed and realized using the CEBAF electron beam and the JLab Hall A equipment.

\subsection{Apparatus} \label{subsec-setup}

Since the Hall was in its commissioning phase at the time of the data taking for this experiment (1998), not all devices were fully operational and the minimal number of detectors were used. However the experiment fully exploited the main capabilities of the accelerator and the basic Hall equipment: 100\% duty cycle, high resolution spectrometers, high luminosities. The Hall A instrumentation is described extensively in ref.~\cite{Alcorn:2003} and in several thesis works related to the experiment~\cite{Degrande:2001th,Jaminion:2001th,Jutier:2001th,Laveissiere:2001th,Todor:2000th}. Only a short overview is given here, and some specific details are given in the subsections.


The continuous electron beam at 4 GeV energy (unpolarized) was sent to a 15 cm long liquid hydrogen target. The two High Resolution Spectrometers, noted here \hrse \ and \hrshnospace, were used to detect in coincidence an outgoing electron and proton, respectively. After exiting the target region the particles in each HRS encounter successively: the entrance collimator of 6 msr, the  magnetic system (QQDQ), and the detector package. The latter consisted of a set of four vertical drift chambers (VDC), followed by two scintillator planes S1 and S2. It was complemented in the \hrse by a Cerenkov detector and a shower counter, and in the \hrsh by a focal plane polarimeter. The VDCs provided particle tracking in the focal plane. The scintillators were the main trigger elements. They provided the timing information in each spectrometer and allowed to form the coincidence trigger.

\subsection{Kinematical settings and data taking} \label{subsec-kin}

\begin{figure}[htb]
\includegraphics[width=8.3cm]{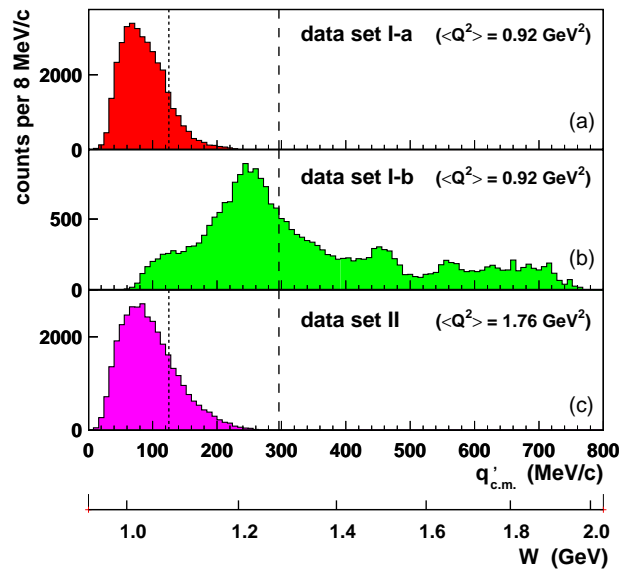}  
\caption{(Color online) The range in $q'_{c.m.}$, or $W$, covered by the various data sets for the $\epg$ events. The vertical lines show the upper limit applied in the analyses:  the pion threshold (dotted line at $W=1.073$ GeV) for the LEX analyses and $W=1.28$ GeV (dashed line) for the DR  analyses. $W$ and $q'_{c.m.}$ are related by \ $W= q'_{c.m.} + \sqrt{ q'^2_{c.m.} + \nucleonmass ^2 }$. 
}
\label{fig-rangeinw}
\end{figure}

Data were taken in two different $Q^2$-ranges, near 0.9 and 1.8 GeV$^2$. The corresponding data sets are labelled I and II, respectively. At $Q^2=0.9$ GeV$^2$ dedicated data were taken in the region of the nucleon resonances~\cite{Laveissiere:2008zn}. This leads us to split data set I into two independent subsets, I-a and I-b, according to the $W$-range.  Figure~\ref{fig-rangeinw} displays the various domains covered in $W$, or equivalently $q'_{c.m.}$. Data sets I-a and II have events essentially below the pion threshold, while data set I-b is more focused on the $\Delta(1232)$ resonance region and above. For the analyses presented here, emphasis will be put on data sets I-a and II. For data set I-b, details can be found already in \cite{Laveissiere:2008zn}, in which a nucleon resonance study was performed up to $W=2$ GeV. Here, the lowest-$W$ part of data set I-b is analyzed in terms of GPs. Table~\ref{tab-data-sets} summarizes our notations.

\begin{table}[htb]
\caption{\label{tab-data-sets} The various data sets of the experiment. Columns 2 and 3 give the ranges in $Q^2$ and $W$ covered by the experiment. The fixed value of $Q^2$ chosen in the analyses is 0.92 GeV$^2$ (resp. 1.76 GeV$^2$) for data sets I-a and I-b (resp. II). Columns 4 and 5 give the $W$-range used in the LEX and DR analyses.
 }
\begin{ruledtabular}
\begin{tabular}{ccccc} 
data & $Q^2$-range &  $W$-range &  $W$-range  &  $W$-range   \\
set  & (GeV$^2$) & (GeV)  & LEX (GeV)  & DR (GeV) \\
\hline
I-a & [0.85, 1.15] &  [0.94, 1.25] & [0.94, 1.07]  & [0.94, 1.25]   \\
\hline
I-b & [0.85, 1.15] &  [1.00, 2.00] &   -----       & [1.00, 1.28]   \\
\hline
II  & [1.60, 2.10] &  [0.94, 1.25] & [0.94, 1.07]  & [0.94, 1.25]   \\
\end{tabular}
\end{ruledtabular}
\end{table}

\begin{figure}[htb]
\includegraphics[width=8.3cm]{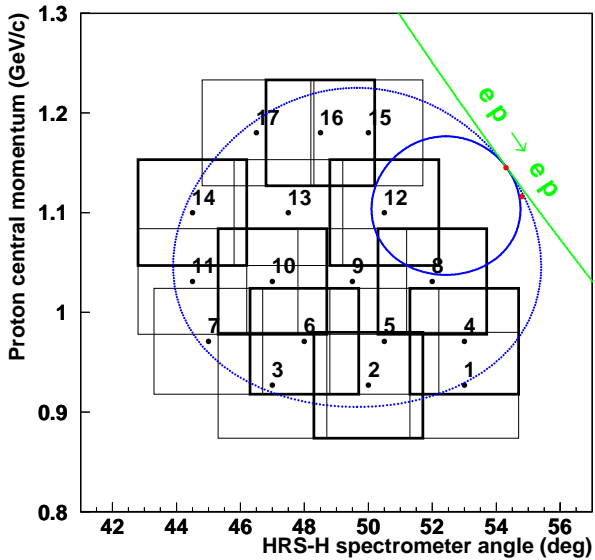}  
\caption{(Color online) The seventeen \hrsh settings for the proton detection in data set I-a. Each setting is represented by a box in momentum and angle. The closed curves correspond to in-plane $\epg$ kinematics at fixed values of $q'_{c.m.}$: 45 MeV/c (inner curve) and 105 MeV/c (outer curve). The \ $ep \to ep$ \ elastic line is also drawn at a beam energy of 4.045 GeV.
}
\label{fig-settingsda1}
\end{figure}

For each of the data sets I-a and II, the \hrse setting was kept fixed, while the \hrsh setting was varied in momentum and angle. In process (1) the final proton is emitted in the lab system inside a cone of a few degrees around the direction of the virtual photon, thanks to a strong CM-to-Lab Lorentz boost. Therefore with a limited number of settings (and in-plane spectrometers),  one can cover most of the desired  phase space, including the most out-of-plane angles. As an example,  Fig.~\ref{fig-settingsda1} illustrates the configuration of the \hrsh settings for data set I-a. In addition, in the \hrse the momentum setting is chosen in order to have the VCS events in the center of the acceptance, i.e. near $\delta p/p=0$\%. As a result, the elastic peak from  \ $ep \to ep$ \ scattering may also be in the acceptance of this spectrometer (at higher $\delta p/p$), especially when $W$ is low, i.e. for data sets I-a and II. In this case, electrons elastically scattered from hydrogen are seen in the \hrse single-arm events, although they are kinematically excluded from the true coincidences.

\begin{table}[hb]
\caption{\label{tab-settings} Summary of the kinematical settings for each data set.
The nominal beam energy is $E_{beam}=$ 4.045 GeV (see section~\ref{subsec-calib} for actual values). $p_e$ and $\theta_e$ are the central momentum of the \hrse
spectrometer and its angle w.r.t. the beamline (electron side). $p_p$ and $\theta_p$ are the same variables on the proton side, i.e. for the \hrsh spectrometer.
}
\begin{ruledtabular}
\begin{tabular}{ccccc} 
data & $p_e$ & $\theta_e$ & $p_p$ & $\theta_p$   \\
set  &       (GeV/c)  & (deg)  & (GeV/c) & (deg) \\
\hline
I-a & 3.43          & 15.4          & [0.93, 1.19]    & [45, 53]  \\

I-b & [3.03, 3.26]  & [15.7, 16.4]  & [1.31, 1.53]    & [37, 45]  \\
II  & 2.93          & 23.0          & [1.42, 1.76]    & [37, 42]  \\
\end{tabular}
\end{ruledtabular}
\end{table}

Data acquisition was performed with the CODA software developed by CEBAF~\cite{CODA}. The trigger setup includes several types, among which T1 and T3 are single-arm \hrse and \hrsh good triggers. The T5 triggers, formed by the coincidence between T1 and T3, are the main ones used in the physics analysis. For each event the raw information from the detectors and the beam position devices is written on file. Scalers containing trigger rates and integrated beam charge are inserted periodically in the datastream, as well as various parameters from the EPICS slow control system. Special runs were recorded to study spectrometer optics.


\section{Data analysis} \label{sect-data-ana}

This section describes the main steps that were necessary to reach the accurate measurement of the $( ep \to ep \gamma)$ differential cross section: raw-level processing, event reconstruction, analysis cuts, and acceptance calculation.

\subsection{Beam charge, target density and luminosity} \label{subsec-lumi}


The electron beam current is measured by two resonant cavities (Beam Charge Monitors BCM) placed upstream of the Hall A target. The signal of each cavity is sent to different electronic chains. In the experiment, the main measurement of the  beam charge used the upstream cavity and the chain consisting in an RMS-to-DC converter followed by a Voltage-to-Frequency converter (VtoF), generating pulses that are counted by a scaler. The content of the VtoF scaler was written on the runfile every 10 seconds. At the end of each run one obtained in this way the integrated charge of the run. The BCM were calibrated twice a day against the Unser monitor, located between the two cavities and measuring the beam current in absolute. The procedure also implied the offline calibration of the VtoF converter. Beam currents ranged from 30 to 100 $\mu A$ with an average of 70 $\mu A$, and the integrated charge per run was determined with an accuracy of 0.5\%.


The experiment used the 15 cm long liquid hydrogen cell (``beer can''). It was kept at a constant temperature $T= 19.0$ K and pressure $P=1.725$ bar, yielding a density $\rho_0=0.0723$ g.cm$^{-3}$ at zero beam current \cite{Suleiman:1998}. The beam was rastered on the target in both transverse directions in order to avoid local boiling of the hydrogen. Studies based on the data of this experiment~\cite{Jutier:2001th} showed that density losses reached at maximum 1\% for a beam current of 100 $\mu A$, so the target boiling effect was considered to be negligible and the density was taken equal to $\rho_0$ in the analyses.


The luminosity $\mathcal{L}$ needed for cross section measurements is obtained on a run-per-run basis. Based on the above considerations, it is determined with an accuracy of $\sim$ $\pm$ 1\%. Typical values of instantaneous luminosities are of the order of 2 to 4$\times 10^{38}$ cm$^{-2} \cdot$s$^{-1}$.

\subsection{Rate corrections} \label{subsec-rate}

The raw event rate is obtained  by counting the number of T5 events, i.e. the coincidence triggers between the two spectrometers.
Several correction factors have to be applied to this rate.

The first correction is due to trigger inefficiency, coming from the scintillators of the detector package. It is obtained by studying the single-arm ``junk triggers'' T2 (electron side) and T4 (hadron side), which record all configurations other than normal in the scintillators. The normal configuration (T3 or T5) is a coincidence between paddles in the S1 and S2 planes in an allowed geometrical configuration (``S-ray''), each paddle signal requiring the coincidence between its left and right phototubes. Among the junk triggers, there are some good events, typically with a hit missing in the scintillator paddles. We identify them by a ``clean-up'' procedure, consisting in the additional requirement of a valid track in the VDCs and a Cerenkov signal in the electron arm. The scintillator inefficiency is then defined as the number of such good T2 or T4 events, relative to the number of (T1+T5) or (T3+T5) events in the same clean-up conditions. The inefficiency is calculated independently for both planes S1 and S2 in each arm. It is binned in the $x$ (dispersive) and $y$ (transverse) coordinates in each plane, to account for local variations. The observed inefficiency  was usually of the order of one percent, reaching occasionally  10\% locally in the electron arm \cite{rds:2001}. This commissioning problem was fixed after our experiment.

The second correction concerns the acquisition deadtime. For each run, a scaler counts the number  $S_5$ of T5 events at the output of the trigger logic. Among these events, only $N_5$ ($< S_5$) are  written on file due to the deadtime of the [acquisition+computer] system. The correction consists in multiplying the event rate by the ratio $S_5/N_5$. The deadtime depends on the beam current; it varied between 5\% and 40\% in the experiment.

The third correction comes from the deadtime of the trigger electronics itself (EDT). It was not measured directly during the experiment but determined afterwards. The EDT estimation is based on a fit to the actual rates in the scintillator paddles, using the strobe rate of each spectrometer. This fit was established in later experiments when the strobe rate was inserted in the datastream~\cite{EDT2000}. The resulting correction is of the order of 1-3\% in our case.

The tracking inefficiency is considered to be negligible, in the sense that for a real particle, the tracking algorithm basically always finds at least one track in the focal plane, which allows to process the event further. This is due to the good efficiency per wire in the VDCs.

Finally, another small correction of the order of 1\% is applied to account for the losses of protons by nuclear interactions in the target and spectrometer windows.

%

The uncertainty on the event rate, after having corrected for all the above inefficiencies, is estimated to be smaller than 0.5\% in relative.

\subsection{Event reconstruction} \label{subsec-recons}

\begin{figure}[htb]
\includegraphics[width=8.3cm]{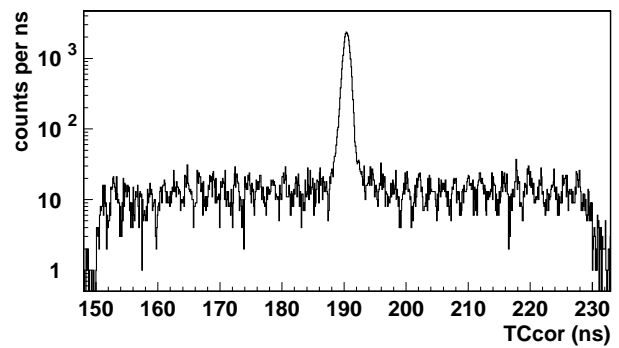}  
\caption{Coincidence time spectrum of data set I-a. The central peak is 0.5 ns wide in rms. 
}
\label{fig-tccor}
\end{figure}

The Hall A analyzer ESPACE processes the raw detector signals and builds all the first-level variables of the event: coincidence time, beam position on target, three-momentum vector of the detected particles at the vertex point, etc.  Figure~\ref{fig-tccor} shows a typical coincidence time spectrum between the \hrse and \hrshnospace. The central peak allows to select the true coincidences. Random coincidences under the peak are subtracted using side bands. In the plateau one clearly sees the 2 ns microstructure of the beam, corresponding to Hall A receiving one third of the pulses of the 1497 GHz CW beam.

In the analyses presented here, particle identification in the detectors is basically not needed. This is because the kinematical settings of VCS near threshold (cf. Table~\ref{tab-settings}) are close to $ep$  elastic scattering, therefore the true coincidences between the two spectrometers are essentially  $(e,p)$ ones. The other true coincidences that could be considered are of the type $(e^-, \pi^+)$ or $(\pi^-, p)$, coming from single or multiple pion production processes. However, such events either do not match the acceptance settings (case of single charged-pion electroproduction) or they yield missing masses which are beyond one pion mass, i.e. far from the VCS region of interest (case of multiple pion production). Therefore detectors such as the gas Cerenkov counter or the electromagnetic calorimeter in the \hrse were essentially not used, and only the information from the VDCs and the scintillators in both arms were treated. As a verification, however, the analysis of data set II was performed with and without requiring a signal in the Cerenkov counter in the \hrsenospace, and the results were found unchanged.


Due to the extended size of the target and the rather large raster size ($\sim \pm$ 3-5  mm in both directions), it is important to know the interaction point inside the hydrogen volume for each event. This point is characterized by its coordinates ($\xvertex, \yvertex, \zvertex$) in the Hall A laboratory frame. The coordinates transverse to the beam axis, horizontal $\xvertex$ and vertical $\yvertex$,  are obtained from the two BPMs located upstream the target. It turned out that for a large fraction of the data taking, the BPM information was accidentally desynchronized from the event recorded by CODA. A special re-synchronization procedure  \cite{Hyde:2001,Laveissiere:2001th} was established offline by coupling the BPM to the raster information (which is always synchronized with the event). Then the BPMs could be used, yielding  $\xvertex$ and $\yvertex$ in absolute to better than $\pm$ 0.5 mm event per event.

The calculation of the longitudinal coordinate $\zvertex$ requires information from the spectrometers. It is obtained by intersecting the beam direction with the track of one of the two detected particles. For this task the \hrse was chosen, since it has the best resolution in horizontal coordinate, i.e. the variable noted $y_{tg}$ in the spectrometer frame. The resolution in $y_{tg}$ is excellent, about 0.6 mm in rms for the \hrse (and twice larger for the \hrshnospace).


The particle reconstruction proceeds as follows. In each arm the particle's trajectory is given by the VDCs in the focal plane. This ``golden track'' is characterized by four independent variables $(x,y,\theta,\varphi )_{fp}$. These variables are combined with the optic tensor of the spectrometer to yield four independent variables at the target: the relative momentum $\delta p/p$, the horizontal coordinate $y_{tg}$, and the projected horizontal and vertical angles,  $\theta_{tg}$ and $\varphi_{tg}$, in the spectrometer frame. A fifth variable $x_{tg}$ characterizing the vertical extension of the track is calculated using the beam vertical position, and allows to compute  small extended-target corrections to the dispersive variables $\theta_{tg}$ and $\delta p/p$. The total energy of the particle is then determined from its momentum and its assumed nature ($e$ in \hrse or $p$ in \hrshnospace). It is further corrected for energy losses in all materials, from the interaction point to the spectrometer entrance. At this level the four-momentum vectors of the incoming electron, scattered electron and outgoing proton:  $ k, k'$  and $p'$, respectively, are known at the vertex point. One can then compute the full kinematics of the reaction $ ep \to ep X$ and a number of second-level variables.

For the physics analyses, the main reconstructed variables are $Q^2, W, \epsilon, q_{c.m.}$ on the lepton side, and the four-momentum vector of the missing system $X$: $p_X = k+p-k'-p'$. This four-vector is transformed from the laboratory frame to the \CMnospace, where one calculates the angles of the missing momentum vector $\vec p_{Xc.m.}$ w.r.t. the virtual photon momentum vector $\vec q_{c.m.}$. These angles, polar $\theta_{c.m.}$ and azimuthal $\varphi$ (see  Fig.~\ref{fig-epgamma-kinem}), and the modulus of the missing momentum, which represents $q'_{c.m.}$ in the case of VCS, are the three variables used to bin the cross section.

Other second-level variables are important for the event selection as well as for the experimental calibration. The first one is the missing mass squared $M_X^2 = (k+p-k'-p')^2$. In the experimental $M_X^2$ spectrum, a photon peak and a $\pi^0$ peak are observed, corresponding to the physical processes $ep \to ep \gamma$  and $ep \to ep \pi^0$ (see  Fig.~\ref{fig-rawmx2}). A cut in $M_X^2$ is thus necessary to select the reaction channel.

Two other variables, of geometrical nature, have proven to be useful. The first one, $\ddd$, compares two independent determinations of the horizontal coordinate of the vertex point in the Hall A laboratory frame: the one measured by the BPMs ($\xvertex$) and the one obtained by intersecting the two tracks measured by the spectrometers, and called  $\twoarmx$. The distribution of the difference $\ddd = \twoarmx - \xvertex$ is expected to be a narrow peak centered on zero. The second geometrical variable, $Y_{dif}$, will be described in section~\ref{subsec-anacuts}.

\subsection{Experimental calibration} \label{subsec-calib}

A lot of experimental parameters have to be well calibrated. At the time of the experiment, the existing optic tensors of the spectrometers were not fully adapted to an extended target; it was necessary to optimize them. Using dedicated runs, new optic tensors were determined for the VCS analysis \cite{sja:2001}. They were obtained in both arms for our designed momentum range, and they  clearly improved the resolution of the event reconstruction.

A number of offsets, of either geometrical or kinematical origin, also had to be adjusted. Among the geometrical offsets, some were given by the CEBAF surveys, such as the target and collimator positions. Others, such as the mispointing of the spectrometers, were recorded in the datastream, but their reading was not always reliable and some of them had to be adjusted by software. One should note that not all offsets have to be known in absolute; what is needed is the relative coherence between target position, beam position and spectrometer mispointing. The consistency checks were made on the distribution of the $\zvertex$ and $\ddd$  variables (defined in section~\ref{subsec-recons}) for real events. The main geometrical offset was found to be a horizontal mispointing of the \hrse by 4 mm.

\begin{figure}[htb]
\includegraphics[width=8.3cm]{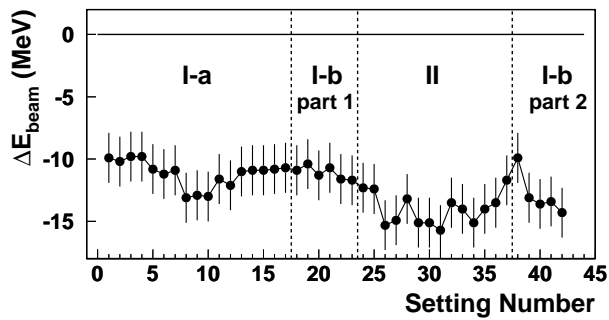}  
\caption{The fitted offset in beam energy, $\Delta E_{beam}$, versus the setting number (time-ordered). There is one point per setting. The various data sets are delimited by the vertical lines. The horizontal line at $\Delta E_{beam}=0$ corresponds to the nominal beam energy from the accelerator, $E_{beam}=4.045$ GeV.
 }
\label{fig-debeam}
\end{figure}

The kinematical offsets consist in small systematic shifts in the reconstructed momentum and angles of the particles at the vertex point. They are mainly due to {\it i)} a beam energy uncertainty (the beam energy measurements described in~\cite{Alcorn:2003} were not yet operational),  {\it ii)}  residual biases in the optic tensors,  {\it iii)} field reproducibility in the spectrometer magnets. The adjustment of these offsets is based on the optimization of the peaks in missing mass squared, in width and position, for the two reactions $ep \to ep \gamma$ and $ ep \to ep \pi^0$ simultaneously. This procedure yields a coherent set of offsets for each setting \cite{hfoff:2001}. An overview of the results is presented in Table~\ref{tab-offsets}. All kinematical offsets were found to be small, except for the beam energy which was significantly below the nominal value from the accelerator, by about 10-16 MeV (see  Fig.~\ref{fig-debeam}).

\begin{table}[ht]
\caption{\label{tab-offsets} Global results for the fitted offsets on the seven variables: beam energy, particle momenta ($p_e, p_p$) and particle angles. $\varphi_{tg(e)}, \varphi_{tg(p)}$ (resp. $\theta_{tg(e)}, \theta_{tg(p)}$)
are the horizontal (resp. vertical) angles of the particle's track in the spectrometer frame. Some offsets have to be fixed in order to ensure the fit stability \cite{hfoff:2001}. The range in brackets indicates setting-to-setting variations of the offsets.
}
\begin{ruledtabular}
\begin{tabular}{ccc} 
variable & range found  & estimated uncertainty \\
\        & for the offset       & on the offset         \\
\hline
$E_{beam}$     & [-16, -10] MeV       & $\pm$ 2 MeV      \\
$p_e$          & 0   \ (fixed)        & $\pm$ 0.3 MeV/c     \\
$p_p$          & [-1.5, +1.5] MeV/c   & $\pm$ 0.5 MeV/c     \\ 
$\varphi_{tg(e)}$    & [0, +0.1] mr         & $\pm$ 0.3 mr     \\ 
$\varphi_{tg(p)}$    & [-1.7, -0.7] mr      & $\pm$ 0.3 mr     \\
$\theta_{tg(e)}$     & [-1.6, -0.5] mr      & $\pm$ 0.5 mr     \\
$\theta_{tg(p)}$     & 0    \ (fixed)       & $\pm$ 0.5 mr     \\
\end{tabular}
\end{ruledtabular}
\end{table}

\subsection{Analysis cuts} \label{subsec-anacuts}

\begin{figure}[htb]
\includegraphics[width=8.3cm]{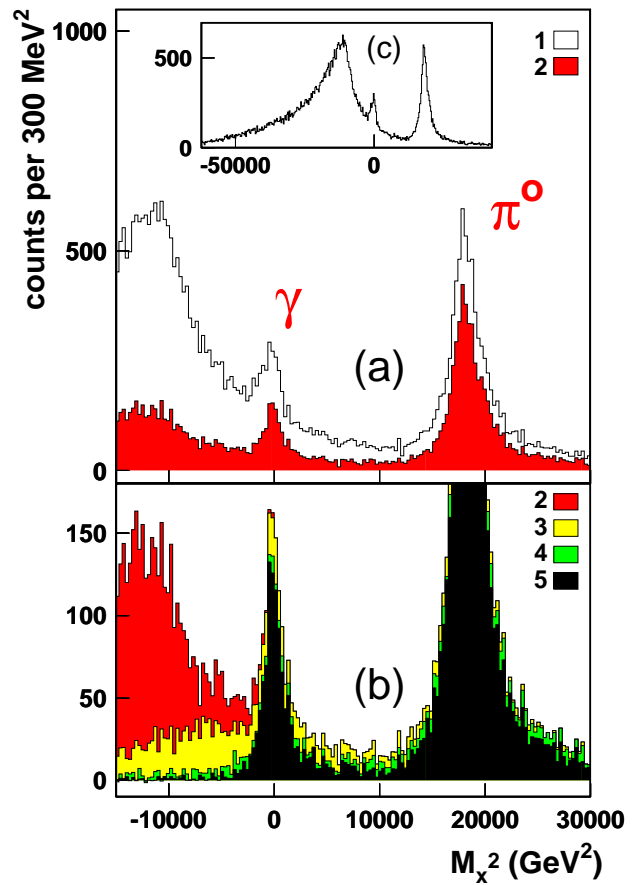}  
\caption{(Color online) A sample of data set II: the experimental spectrum of the missing mass squared at various levels of cuts, added successively and labelled from 1 to 5.  (a): the raw coincidences (1) and adding the R-function cut (2). (b): adding the conditions $W>0.96$ GeV (3),  $Y_{dif}<-0.012$ m (4) and $\vert \ddd \vert < 3$ mm (5) (see the text for the description of the variables). 
The insert (c) shows histogram 1 in full scale abscissa.
}
\label{fig-rawmx2}
\end{figure}

The offsets described above were established using clean event samples. However, the raw coincidences are not so clean, as can be seen e.g. from the spectrum in the insert of  Fig.~\ref{fig-rawmx2} (top plot). The photon peak is contaminated by a large broad bump centered near -15000 MeV$^2$. These events are mostly due to $ep$ elastic scattering with the final proton ``punching through'' the \hrsh entrance collimator. They require specific cuts in order to be eliminated. A key condition for the VCS analysis is to obtain a well-isolated photon peak in the $M_X^2$ spectrum ( Fig.\ref{fig-rawmx2}, histogram 5). The cuts necessary to reach this goal are described below.

%
%
First, standard acceptance cuts are applied in each arm. They use essentially the Hall A R-functions~\cite{Rvachev}, which are a way to handle complex cuts in a multi-dimensional space. In the R-function approach, the problem is transformed into the calculation, for each detected particle, of its "distances" to the acceptance boundaries, and the combination of these distances into one single function. This R-function takes continuous values: positive inside the acceptance domain, negative outside, and equal to zero on the boundaries; it can then be used as a one-dimensional cut (e.g., here we require R-function $>$0). We also use additional -and largely redundant- contour cuts in two dimensions among the ($\delta,y,\theta,\varphi)_{tg}$ quadruplet, and restricted apertures in the plane of the entrance collimators. We note that the target endcaps, located at $\pm$ 75 mm on the beam axis, are not seen in coincidence, due to the rather large \hrsh spectrometer angles; so a cut in $\zvertex$ is not necessary. The effect of the standard acceptance cuts is shown in  Fig.~\ref{fig-rawmx2} (histogram 2). Clearly they are not sufficient to fully clean the $M_X^2$ spectrum, and supplementary cuts are necessary.
%

Normally, the protons coming from $ep$ elastic scattering are too energetic to be in the momentum acceptance of the \hrsh for our chosen settings. However, some of these protons go through the material of the entrance collimator (tungsten of 80 mm thickness) where they scatter and lose energy, after which they enter the acceptance and are detected. This problem cannot be avoided, since VCS near threshold is by nature close to $ep$ elastic scattering.

\begin{figure}[htb]
\includegraphics[width=8.3cm]{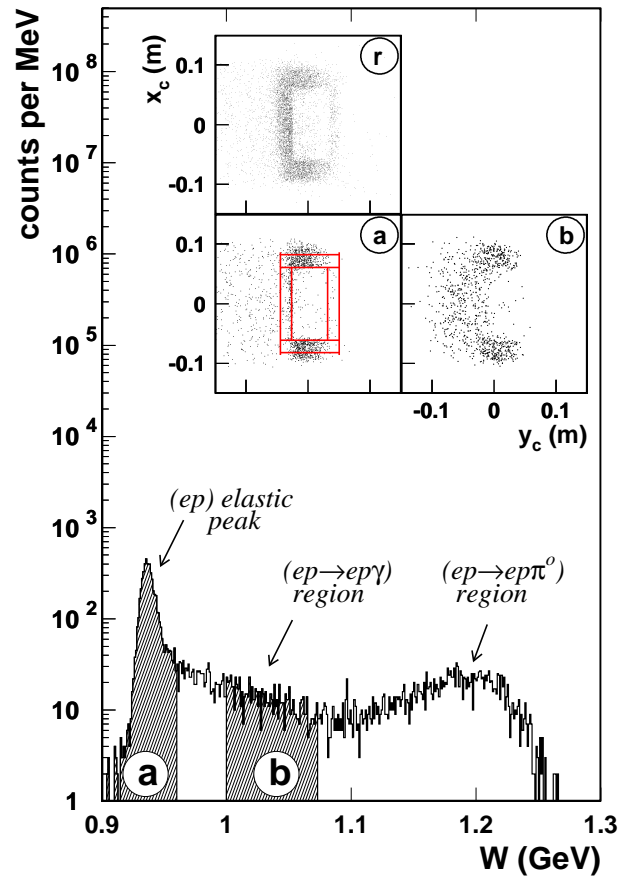}  
\caption{(Color online) A sample of data set II: the experimental  $W$ spectrum for the coincidence events after the R-function cut. The inserts show the proton impact in the \hrsh collimator calculated ``elastically'' (see text). Events in inserts (a) and (b) correspond to the two hatched zones of the histogram: the $ep$ elastic peak ($W<0.96$ GeV (a)) and a typical VCS region ($1.0<W<1.073$ (b)). In insert (a) a sketch of the tungsten collimator is drawn. The upper insert (r) shows the $ep$ elastic events before applying the R-function cut.
 }
\label{fig-punch0}
\end{figure}

As a result, a prominent  $ep$ elastic peak is seen in the $W$-spectrum for true coincidences, at the raw level and even after having applied the standard acceptance cuts, cf.  Fig.~\ref{fig-punch0}. A striking evidence for ``punch-through'' protons is also provided by the inserts in  Fig.~\ref{fig-punch0}. These plots show the 2D impact point of the proton in the \hrsh collimator plane, calculated in a particular way. Here we do not use the information from the \hrshnospace, we only use  the \hrse informations and the two-body $ep$ elastic kinematics. Knowing the vertex point (from the \hrse and the beam), the beam energy and the scattered electron angles, one can deduce the point  where the proton from $ep$ elastic kinematics should have hit the collimator. This hit point is characterized by its coordinates  $x_c$ (vertical) and $y_c$ (horizontal) in the \hrsh spectrometer frame. For events in the elastic peak of  Fig.~\ref{fig-punch0}, the $(x_c,y_c)$ distribution (insert (r)) reproduces faithfully the structure of the collimator material,  proving that these are indeed protons punching through the collimator. The  R-function cut is able to remove part of these events, but keeps the punching through the upper and lower parts of the collimator, as shown  by insert (a). Of course our purpose here is only illustrative; these events are not a concern, since they are removed by a simple cut in $W$ around the elastic peak. The result of such a cut in terms of missing mass squared is displayed as histogram 3 of  Fig.~\ref{fig-rawmx2}.

%
%

The main concern is that there are also ``punch-through'' protons at higher $W$, as evidenced by insert (b) of  Fig.~\ref{fig-punch0}, where an image of the collimator material is still present. This region in $W$ is the far radiative tail of the $ep$ elastic process; in other words it is the kinematical region of interest for VCS, therefore one cannot use a cut in $W$. Nevertheless these ``punch-through'' events must be eliminated, because $i$) they are badly reconstructed and  $ii$) the simulation cannot reproduce them (our simulation, which is used to obtain the cross section, see sections \ref{subsec-solidangle} and \ref{subsec-cross1}, considers only perfectly absorbing collimators). To this aim, a more elaborate cut has been designed which we now describe.

For a ``punch-through'' proton, the variables  ($\delta,y,\theta,\varphi)_{tg}$ obtained directly from the \hrsh are usually severely biased, due to the crossing of a thick collimator. Therefore they are of little use, except for one particular combination:  $\ycollih = y_{tg}+ D \cdot \varphi_{tg}$, where $D$ is the distance from the target center to the collimator. This quantity $\ycollih$ gives the horizontal impact coordinate of the proton track in the collimator plane, as measured directly by the \hrshnospace . It is an unbiased variable, even for a ``punch-through'' proton. This is because the collimator plane is where the distortion happened.  The reconstruction of the proton trajectory, which is performed backwards, from focal plane to target, is correct down to the entrance collimator (and biased further down to the target). The idea is then to compare this quantity $\ycollih$ with the ``elastic'' coordinate $y_c$ calculated just above. For ``punch-through'' protons the two calculations turn out to be in close agreement, hence the difference  $Y_{dif}=y_c - \ycollih$ peaks at zero. We point out that this comparison can only be done for the horizontal coordinate, not in vertical, due to the fact that the vertical extension $x_{tg}$ is intrinsically not measured by the spectrometers.

\begin{figure}[htb]
\includegraphics[width=8.3cm]{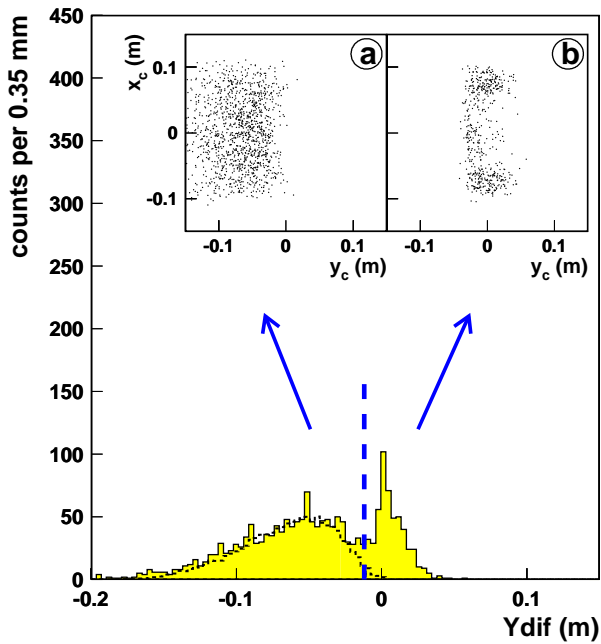}
\caption{(Color online) A sample of data set II: the experimental $Y_{dif}$ spectrum (see text) for coincidence events surviving the AND of the three following cuts: R-function $>0$, $-5000 < M_X^2 < 5000$ MeV$^2$ and $W>0.96$ GeV  (solid histogram). The inserts show the ``elastically calculated'' proton impact in the \hrsh collimator (see text) for $Y_{dif}<-0.012$m (clean events, plot (a)) and for $Y_{dif}>-0.012$ m (punch-through protons, plot (b)). The dashed histogram corresponds to the VCS simulation with the same three cuts.
}
\label{fig-punch2}
\end{figure}

 Fig.~\ref{fig-punch2} shows the $Y_{dif}$ distribution. Clean VCS events cover smoothly the $Y_{dif}<0$  region (dashed histogram), while experimental events (solid histogram) exhibit an extra-peak centered on zero. This peak corresponds to ``punch-through'' protons and is most efficiently eliminated by requiring the condition  $Y_{dif}<-12$ mm.  The rejected events (insert (b)) again clearly reveal the image of the tungsten collimator. The retained events (insert (a)) show a  smooth distribution in the $(x_c,y_c)$ plane, well reproduced by the VCS simulation.  The  $Y_{dif}$ cut is definitely efficient in isolating a clean photon peak, as shown by histogram 4 of  Fig.~\ref{fig-rawmx2}.

Lastly, to obtain histogram 5 in  Fig.~\ref{fig-rawmx2} the geometrical variable $\ddd$ (cf. section~\ref{subsec-recons} and  Fig.~\ref{fig-d-and-mx2}) is selected around the central peak, i.e. in the interval [-3,+3] mm,  completing the removal of badly reconstructed events. It is worth noting that these two last cuts in $Y_{dif}$ and $\ddd$ (which are correlated but not equivalent),  owe their efficiency to the excellent spectrometer intrinsic resolution in $y_{tg}$, already emphasized in section~\ref{subsec-recons}.

After the above cuts, a small fraction of events ($\le$ 5\%) still have more than one track in  one arm or the other. The number of tracks  is given by the VDC algorithm together with  the parameters of the ``golden track''. One may either keep these multi-track events, or reject them and renormalize the rate accordingly, based on the fact that these are still good events, just less clean. This second method was chosen, except for data set II where the multi-track events are in very small proportion ($\le$ 0.5\%). Finally,  events are selected in a window in $M_X^2$ around the photon peak, typically [-5000,+5000] MeV$^2$, and in certain $W$-range. The lower bound in $W$ corresponds to $q'_{c.m.}=30$ MeV/c and the upper bound is imposed depending on the type of analysis, LEX or DR (cf. Fig.\ref{fig-rangeinw}).   The (very few) random coincidences that remain are subtracted. After all cuts, the final event statistics for the analyses are about 35000 (data set I-a), 13000 (data set I-b) and  25000 (data set II).

\subsection{Monte-Carlo simulation} \label{subsec-solidangle}

The experimental acceptance is calculated by a dedicated Monte-Carlo simulation which includes the actual beam configuration, target geometry, spectrometor acceptance and resolution effects. It is described in detail in~\cite{Janssens:2006} and we just recall here the main features. The $\epg$ kinematics are generated by sampling in the five variables of the differential cross section $d^5 \sigma / d k'_{e lab} d \Omega'_{e lab} d\Omega_{\gamma c.m.}$. The scattered electron momentum and angles in the laboratory frame define the virtual photon, then with the angles of the Compton process in the \CM one can build the complete 3-body kinematics. Events are generated according to a realistic cross section, the (BH+Born) one, over a very large phase space. The emitted electron and proton are followed in the target materials, and the event is kept if both particles go through the [collimator+spectrometer] acceptance. One forms the four target variables ($\delta,y,\theta,\varphi)_{tg}$ of each track  and implements measurement errors on these variables in order to reproduce the resolution observed experimentally. Finally, one proceeds to the event reconstruction and analysis cuts in a way similar to the experiment.

\begin{figure}[htb]
\includegraphics[width=8.3cm]{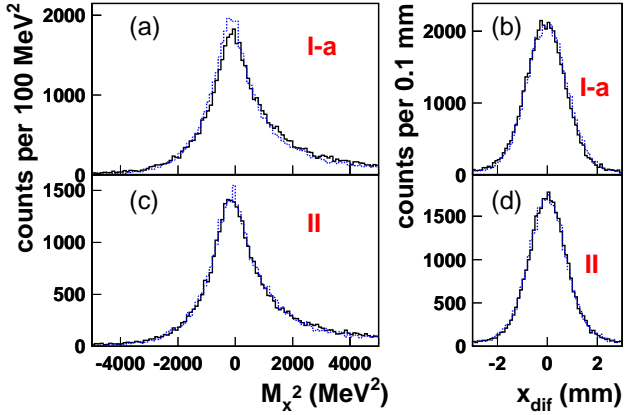}
\caption{(Color online) Data sets I-a (top) and II (bottom) after all cuts: comparison of experiment (solid histogram)  and simulation (dotted histogram). (a),(c): the missing mass squared in the VCS region; the peak FWHM is about 1650 MeV$^2$. (b),(d): the geometrical variable $\ddd$ (see text); the peak FWHM is about 1.9 mm.
}
\label{fig-d-and-mx2}
\end{figure}

As an example,  Fig.~\ref{fig-d-and-mx2} shows the distribution of the variables $M_X^2$ and $\ddd$, for two VCS data sets after all cuts. These variables are not sensitive to the details of the physics; with an infinitely good resolution they should be delta-functions, therefore they characterize the resolution achieved in the experiment. The agreement between the experimental and simulated data is very good not only in the main peak  but also in the tails of the distributions, which is of importance as far as cuts are concerned. The excellent resolution achieved in missing mass squared allows to cleanly separate the $(ep \to ep \gamma)$  and $(ep \to ep \pi^0)$ channels. The residual contamination of $\pi^0$ events under the $\gamma$ peak is negligible for the settings analyzed here: simulation studies show that it is smaller than 0.5 \%.

The radiative corrections are performed along the lines of ref.~\cite{Vanderhaeghen:2000ws}, based on the exponentiation method. The simulation takes into account the internal and external bremsstrahlung of the electrons, because the associated correction depends on the acceptance and the analysis cuts. This allows the simulation to produce a realistic radiative tail in the $M_X^2$ spectrum, visible on the right side of the peak in  Fig.~\ref{fig-d-and-mx2}-left. The remaining part of the radiative effects, due to virtual corrections plus real corrections which do not depend on experimental cuts, is calculated analytically. It is found to be almost constant for the kinematics of the experiment  \cite{hfrad:2000}: $F_{rad} \simeq 0.93$, therefore it is applied as a single numerical factor, such that \  $ d \sigma_{corrected} = d \sigma_{raw} \times F_{rad}$ \ in each physics bin. The estimated uncertainty on $F_{rad}$ is of the order of $\pm 0.02$, i.e. it induces  a $\pm$2\% uncertainty on the cross section, globally on each point and with the same sign.

\subsection{Cross section determination} 
\label{subsec-cross1}

 We first explain the principle of the cross section determination in a bin, and then the chosen binning in phase space. In a given bin, after all cuts and corrections to the event rate, the analysis yields a number of experimental VCS events $N_{exp}$ corresponding to a luminosity $  \mathcal{L}_{exp}$. Similarly, the simulation described in section \ref{subsec-solidangle} yields a number of events $N_{sim}$ corresponding to a luminosity $\mathcal{L}_{sim}$. The experimental cross section is then obtained by:
\begin{eqnarray}
 d^5 \sigma ^{EXP} &=& {N_{exp} \over \mathcal{L}_{exp} } \cdot
{ \mathcal{L}_{sim} \over N_{sim}} \cdot d^5 \sigma_{sim} (P_0) \ , 
\label{eq-dsigma-exp} 
\end{eqnarray} 
where the factor \ $ [ \mathcal{L}_{sim} \cdot d^5 \sigma_{sim} (P_0) / N_{sim}  ] ^{-1}$ \ can be seen as an effective solid angle, or acceptance, computed by the Monte-Carlo method. $ d^5 \sigma_{sim} (P_0)$ is the cross section used in the simulation, at a fixed point $P_0$ that can be chosen freely. As explained in~\cite{Janssens:2006}, this method is justified when the shape of the cross section $ d^5 \sigma_{sim}$  is realistic enough, and it gives rise to a measured cross section ($d^5 \sigma / d k'_{e lab} d \Omega'_{e lab} d\Omega_{\gamma c.m.}$) at some well-defined fixed points in phase space.

\begin{figure}[htb]
\includegraphics[width=8.3cm]{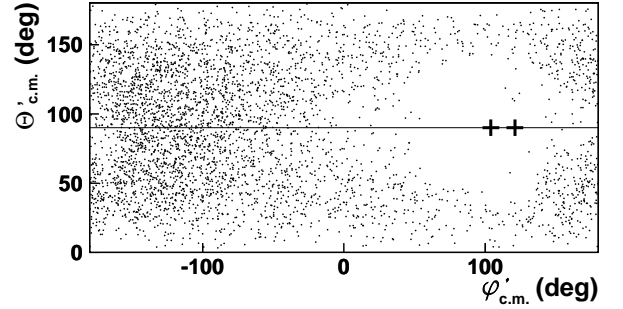}
\caption{
Accepted phase space in ($\theta'_{c.m.}, \varphi'_{c.m.})$ for  data set I-a. The two crosses denote the position of the BH peaks and the horizontal line corresponds to in-plane kinematics ($\theta'_{c.m.}=90^{\circ}$ or $\varphi = 0^{\circ}$ and $180^{\circ}$).
}
\label{fig-theacphiac}
\end{figure}

These points are defined by five independent variables. The most convenient choice w.r.t. the LET formulation is the  set   $(q_{c.m.}, q'_{c.m.}, \epsilon, \theta_{c.m.}, \varphi)$. We will work at fixed $q_{c.m.}$ and fixed $\epsilon$, and make bins in the other three variables.  For the subsequent analyses, instead of the standard ($\theta_{c.m.}, \varphi$) angles, another convention ($\theta'_{c.m.}, \varphi'_{c.m.}$) is chosen. It is deduced from the standard one by a simple rotation: the polar axis for $\theta'_{c.m.}$ is chosen perpendicular to the lepton plane, instead of being aligned with the $\vec q_{c.m.}$ vector (see Appendix \ref{app-1} for more details). This new system of axis allows an angular binning in which the direction of $\vec q _{c.m.}$ does not play a privileged role. Due to the narrow proton cone in the laboratory, the angular acceptance in the \CM  is almost complete for data sets I-a and II. This is illustrated in  Fig.~\ref{fig-theacphiac}. We note that the two peaks of the BH cross section, located in-plane, are out of the acceptance (see also  Fig.~\ref{fig-ratio-1}). This is on purpose, since in these peaks the polarizability effect in the cross section vanishes.  When $W$ increases, the acceptance reduces to more backward scattering angles \cite{Laveissiere:2008zn}.

Table~\ref{tab-bins} in Appendix \ref{app-1} summarizes the bin sizes and the chosen fixed points in phase space. As a consequence, the results of the experiment are obtained at two fixed values of  $Q^2$: 0.92 and 1.76 GeV$^2$.

In eq.(\ref{eq-dsigma-exp}) the cross section $d^5 \sigma ^{EXP}$ is first calculated using  the BH+Born cross section for $d^5 \sigma_{sim}$, i.e. no polarizability effect is included in the simulation. Then, to improve the accuracy, we include a Non-Born term in  $d^5 \sigma_{sim}$, based on what we find for the polarizabilities at the previous iteration. Below the pion threshold this Non-Born term is the first-order LEX term of eq.(\ref{eq-let1}). For the region above the pion threshold,  this Non-Born term is computed using the dispersion relation formalism, and the iterations are made on the free parameters of the model. In all cases this iterative procedure shows good convergence.

\subsection{Sources of systematic errors} 
\label{subsec-systerr-cs}

The systematic errors on the cross section come from three main sources: 1) overall absolute normalization, 2) beam energy, 3) horizontal angles of the detected particles. 

The uncertainty in the absolute normalization has principally three origins: the radiative corrections known to $\pm 2$\%, the experimental luminosity known to  $\pm 1$\%, and the detector efficiency corrections known to  $\pm 0.5$\% (see previous sections). Added in quadrature, they give a overall normalization error of  $\pm 2.3$\%, applying to all cross-section points with the same sign.

The uncertainty in the beam energy, deduced from the offsets study of section~\ref{subsec-calib}, is taken equal to $\pm 2$ MeV. The uncertainty in horizontal angles essentially reflects the accuracy of the optic tensors and is taken equal to $\pm 0.5$ mr in each arm. To study the systematic error induced by the beam energy or the horizontal angles, the experimental events are re-analyzed with these parameters changed, one by one separately, by one standard deviation. One obtains in each case a set of modified cross-section data; in certain cases we observe a change of shape of the cross section. One can summarize these effects by saying that error sources 2) and 3) taken together are equivalent to an average systematic uncertainty of $\pm$ 6\% (resp. $\pm$ 7\%) on the cross section, for data set I-a (resp. II). These errors include substantial point-to-point correlations. 

Systematic errors on the physics observables will be discussed in sections \ref{subsec-lex-ana} and \ref{subsec-dr-ana}.

\subsection{Choice of proton form factors} 
\label{subsec-ff}

The proton elastic form factors $G_E^p$ and $G_M^p$ are an important input in an analysis of VCS at low energy. Indeed they are needed to calculate the BH+Born cross section, which is at the basis of the low-energy expansion. Throughout these analyses the form factor parametrization of Brash {\it et al.}~\cite{Brash:2001qq} was chosen. It provided the first fit consistent with the observed departure from one of the ratio $\mu_p G_E^p/G_M^p$ in our-$Q^2$ range~\cite{Jones:1999rz,Gayou:2001qd}.

The VCS structure functions and GPs are always extracted by measuring a {\it deviation} from the BH+Born process --either analytically as in the LEX approach, or in a more complex way as in the DR approach. This statement means that the GP extraction is sensitive to both cross sections, $d^5 \sigma^{EXP}$ and $d^5 \sigma^{BH+Born}$:  a 1\% change on $d^5 \sigma^{EXP}$ has the same impact as a 1\% change on $d^5 \sigma^{BH+Born}$. This last cross section is not known with an infinite accuracy, due to uncertainties on the proton form factors. Therefore a systematic error should be attached to our calculation of $d^5 \sigma^{BH+Born}$. To treat it in a simplified way, we consider that form factor uncertainties are equivalent to a global scale uncertainty of $\pm$ 2\% on $d^5 \sigma^{BH+Born}$. Then,  when dealing with the extraction of the physics observables (sections \ref{subsec-extract-sf} and \ref{subsec-discuss-gp}), this effect can be put instead on  $d^5 \sigma^{EXP}$, i.e. it can be absorbed in the overall normalization uncertainty of the experiment. Consequently, in sections \ref{subsec-extract-sf} and \ref{subsec-discuss-gp}, we will simply enlarge the systematic error due to normalization (source \# 1 in section \ref{subsec-systerr-cs}) from $\pm$ 2.3\% to $\pm$ 3\% (= quadratic sum of 2.3\% and 2\%).


\section{Results and discussion} 
\label{sec-resul}

We first present the results for the photon electroproduction cross section. Then the VCS structure functions and the GPs are presented and discussed. The main results for these observables are contained in Tables \ref{tab-sf-lex}, \ref{tab-sf-dr} and \ref{resultsgp}.

\subsection{The $\epg$ cross section}\label{subsec-cross2}

The experiment described here provides a unique set of data for VCS studies, combining altogether a large angular phase space (including out-of-plane angles), a large domain in CM energy (from the threshold to the Delta resonance) and an access to the high-$Q^2$ region. Our cross-section data are reported in Tables \ref{tab-cs-1} to  \ref{tab-cs-5} of Appendix~\ref{app-3}.

\subsubsection{Angular and energy dependence}
\label{subsec-cross-ang-ener}

\begin{figure}[htb]
\includegraphics[width=8.3cm]{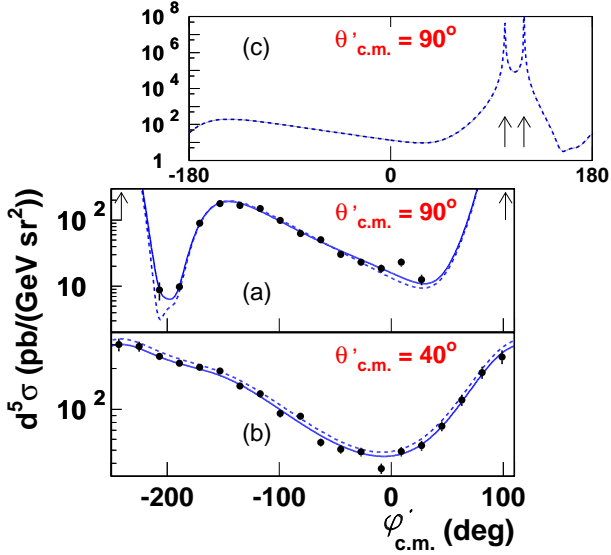}
\caption{(Color online) Data set I-a below the pion threshold, at $q'_{c.m.}=105$ MeV/c. The $(ep \to ep \gamma)$ cross section is shown in-plane ($\theta'_{c.m.}=90^{\circ}$, (a)) and out-of-plane ($\theta'_{c.m.}=40^{\circ}$, (b)). The dotted curve is the BH+Born calculation. The solid curve includes the first-order GP effect calculated using our measured structure functions. The errors on the points are statistical only, as well as in the six next figures. The upper plot (c) shows the in-plane BH+Born cross section with a full-scale ordinate and the more traditional abscissa running between $\varphi'_{c.m.} = -180^\circ$ and  $\varphi'_{c.m.} = +180^\circ$.
}
\label{fig-ratio-1}
\end{figure}

\begin{figure}[htb]
\includegraphics[width=8.3cm]{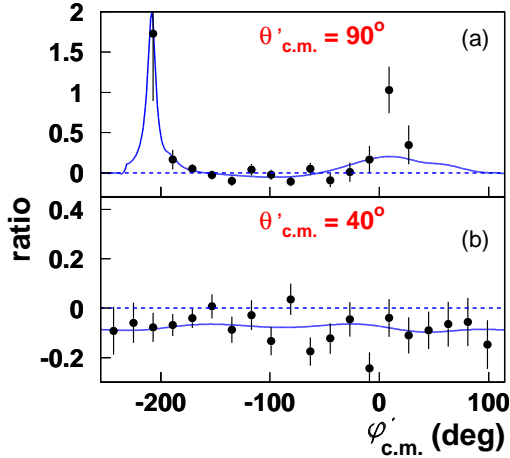}
\caption{(Color online) The ratio $(d \sigma^{EXP} - d \sigma^{BH+Born}) /d \sigma^{BH+Born}$ for the data points of the previous figure. The solid curve shows the first-order GP effect calculated using our measured structure functions. 
}
\label{fig-ratio-2}
\end{figure}

\begin{figure}[htb]
\includegraphics[width=8.3cm]{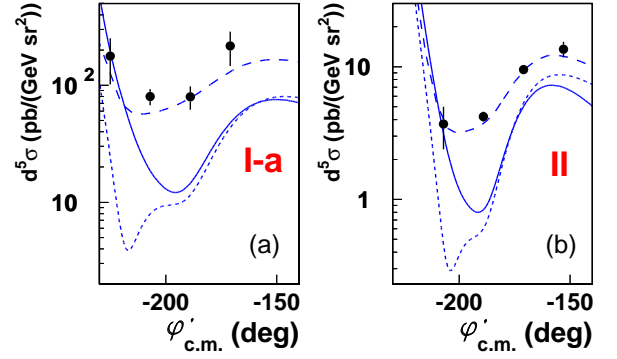}
\caption{(Color online) The $(ep \to ep \gamma)$ cross section for data sets I-a (plot (a)) and II (plot (b)) at $q'_{c.m.}=215$ MeV/c, in-plane ($\theta'_{c.m.}=90^{\circ}$) as a function of $\varphi'_{c.m.}$. The dashed curve is the DR model calculation, with parameter values as fitted in the experiment. The dotted (resp. solid) curve is the BH+Born cross section (resp. plus a first-order GP effect).
}
\label{fig-cs-above}
\end{figure}

Selected samples of our results are presented in Figs. \ref{fig-ratio-1} to \ref{fig-lowqprim}.  Figure~\ref{fig-ratio-1} shows the measured cross section for the highest value of $q'_{c.m.}$ below the pion threshold (105 MeV/c).  The in-plane cross section ($\theta'_{c.m.}=90^{\circ}$) rises by seven orders of magnitude in the vicinity of the BH peaks, which are indicated by the two arrows. The  out-of-plane cross section has a much smoother variation. As expected, the measured values exhibit a slight departure from the BH+Born calculation, due to the polarizabilities. The magnitude of this effect is best seen in  Fig.~\ref{fig-ratio-2} which depicts the deviation of the measured cross section relative to BH+Born: in-plane this ratio varies between -5\% and +20\%,  except in the dip near $\varphi'_{c.m.}=-200^{\circ}$ (or +160$^{\circ}$) where it reaches larger values. This complex pattern is due to the VCS-BH interference. Out-of-plane the polarizability effect is much more uniform, with an average value of $\sim -10$\%.

Another selected sample of results is displayed in  Fig.~\ref{fig-cs-above}, this time above the pion threshold (at $q'_{c.m.}=215$ MeV/c) and for backward angles of the outgoing photon. There, the first-order term of the LET becomes clearly insufficient to explain the observed  cross section, while the calculation of the DR model, which includes all orders, performs quite well. The energy dependence of the cross section, i.e. the dependence in $q'_{c.m.}$ or $W$, is governed by a strong rise when $q'_{c.m.}$ tends to zero due to the vicinity of the $ep$ elastic scattering, and a resonant structure in the region of the Delta(1232). These features can be seen in  Fig.~\ref{fig-secdelta}.

\subsubsection{Overall normalization test} \label{subsec-testnorm}

\begin{figure}[htb]
\includegraphics[width=8.3cm]{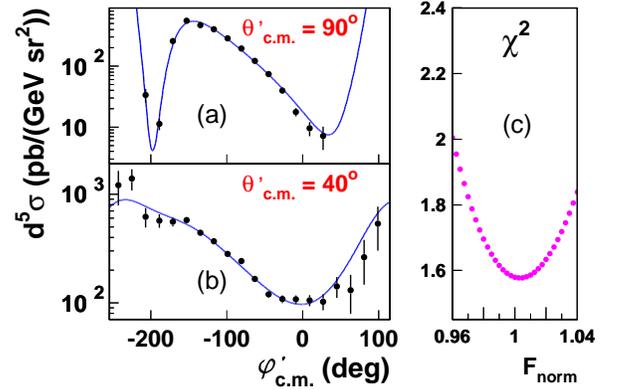}  
\caption{(Color online) Data set I-a. The $(ep \to ep \gamma)$ cross section  at the lowest $q'_{c.m.}$ of 45 MeV/c, for in-plane (a) and out-of-plane (b) kinematics. The solid curve is the (BH+Born + first-order GP) cross section.
The right plot (c) shows the reduced $\chi^2$ of the normalization test.
}
\label{fig-lowqprim}
\end{figure}

The effect of the GPs in the photon electroproduction cross section roughly scales with the outgoing photon energy  $q'_{c.m.}$. Therefore the physics results are determined essentially from the bins at high  $q'_{c.m.}$, which have the highest sensitivity to the GPs. At our lowest $q'_{c.m.}$  of 45 MeV/c, this sensitivity is much reduced, and one can test another  aspect of the experiment, namely the overall normalization.

When $q'_{c.m.}$ tends to zero, $d^5 \sigma^{EXP}$  formally tends to the known BH+Born cross section. This is a model-independent statement, best illustrated by the LEX expansion of eq.(\ref{eq-let1}). At  $q'_{c.m.}=45$ MeV/c, the first-order term ($q'_{c.m.} \cdot \phi \cdot \Psi_0$), calculated using our measured values for the structure functions, is very small. It is  about 2\% of the BH+Born cross section, and it remains essentially unchanged when the structure functions are varied by one standard deviation. Therefore, at the lowest $q'_{c.m.}$ the comparison of the measured cross section  $d \sigma^{EXP}$ with the cross section $d^5 \sigma^{calc}$ calculated from eq.(\ref{eq-let1}):
\begin{eqnarray*}
d^5 \sigma^{calc} =  d^5 \sigma ^{BH+Born}  \ + \ q'_{c.m.} \cdot \phi \cdot \Psi_0 \ , 
\end{eqnarray*}
is essentially a test of the absolute normalization of the experiment. In practice, one allows $d \sigma^{EXP}$ to be renormalized by a free factor $F_{norm}$, and  a $\chi^2$ is minimized  between $d \sigma^{EXP}$  and $d \sigma^{calc}$ as a function of $F_{norm}$. The test is performed on data sets I-a and II at the lowest $q'_{c.m.}$;  the $\chi^2_{min}$ is always found for $F_{norm}$ in the range [0.99, 1.01]. An example is given in  Fig.~\ref{fig-lowqprim}. To conclude, our cross-section data need very little renormalization, less than 1\%. This means in particular that there is a good consistency between the chosen parametrization of the proton form factors, and the way the radiative corrections are applied to the experiment.


\subsection{The VCS structure functions}
\label{subsec-extract-sf}

As mentioned in section \ref{sec-theo}, the VCS structure functions and the GPs do not enter the $\epg$ cross section in the most straightforward way. A theoretical tool is needed to extract them from the experiment. The structure functions have been extracted by two different methods: the LEX analysis  and the DR analysis. This section presents the methods, the results and a discussion.

\subsubsection{LEX analysis} 
\label{subsec-lex-ana}

This analysis is based on the method described in section~\ref{subsec-lex-theo}. It is performed on the data sets I-a and II separately. An upper cut in $W$ is imposed ($W < (\nucleonmass  + \pionmass )$, or $q'_{c.m.}<126$ MeV/c) to stay below the pion threshold. The LEX analysis  uses the cross-section data of Tables \ref{tab-cs-1} and \ref{tab-cs-2} only.

\begin{figure}[htb]
\includegraphics[width=8.3cm]{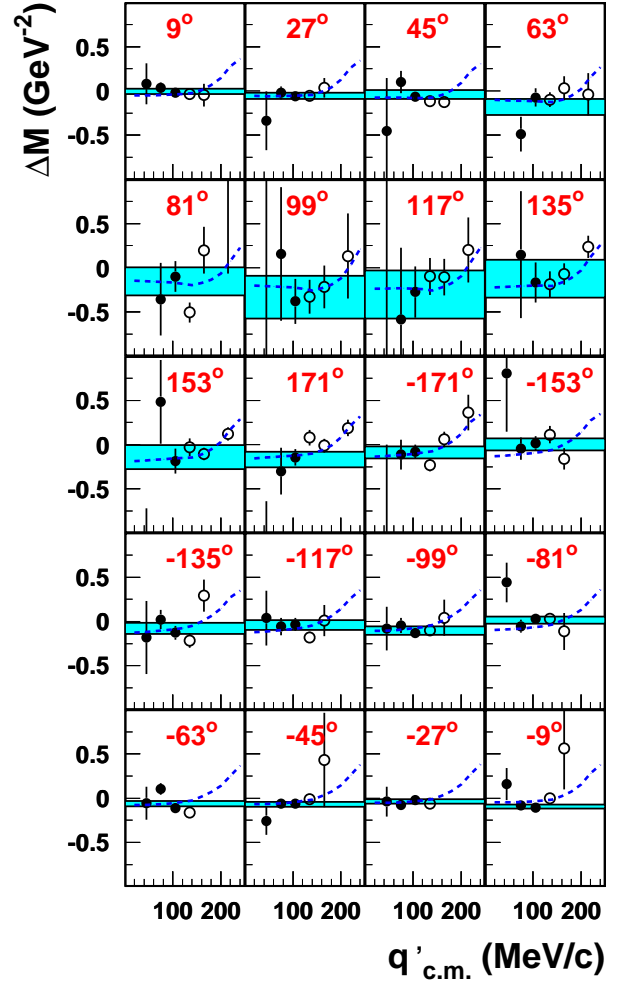}
\caption{(Color online) The $q'_{c.m.}$-dependence of the quantity  $\Delta {\cal M}$ (see eq.(\ref{eq-let3})) in each angular bin. Example of data set I-a at $\theta'_{c.m.}=40^{\circ}$ (out-of-plane). The value of $\varphi'_{c.m.}$ is written in each plot. The shaded band is the result of the LEX fit, ``$\Delta M$=constant'', within $\pm 1 \sigma$ error (statistical). The fit is performed on the three points below the pion threshold (filled circles). The other points (open circles) are above the pion threshold and do not participate to the LEX fit. The dashed curve shows the calculation of  $\Delta {\cal M}$ by the DR model, using the results of our DR fit (see section \ref{subsec-dr-ana}) which is performed on all points (filled + open circles).
}
\label{fig-deltam-lex}
\end{figure}

For each measured point in $(q'_{c.m.}, \theta'_{c.m.}, \varphi'_{c.m.})$,  one forms the quantity:
\begin{eqnarray} \Delta {\cal M} = 
 (d^5 \sigma ^{EXP} -  d^5 \sigma ^{BH+Born}) \, / \, ( q'_{c.m.} \cdot \phi ) \ .
\label{eq-let3} 
\end{eqnarray} 
The  $\Psi_0$ term  of eq.(\ref{eq-let1}) is the extrapolation of $\Delta {\cal M}$ to $q'_{c.m.}=0$ in each bin in $(\theta'_{c.m.}, \varphi'_{c.m.})$. Below the pion threshold, our  $\Delta {\cal M}$ data do not exhibit any significant $q'_{c.m.}$-dependence within error bars. An example is shown in  Fig.~\ref{fig-deltam-lex}. The extrapolation to $q'_{c.m.}=0$ is thus done simply by averaging $\Delta {\cal M}$ over the points at $q'_{c.m.}= $ 45, 75 and 105 MeV/c. This is equivalent to  neglecting the higher-order terms $O(q'^2_{c.m.})$ in this $q'_{c.m.}$-range.

\begin{figure}[htb]
\includegraphics[width=8.3cm]{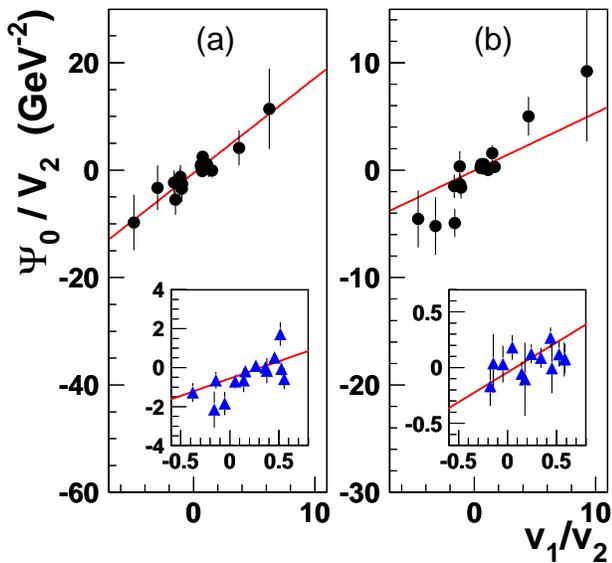}
\caption{(Color online) A graphical representation of the LEX fit for data sets I-a (plot (a)) and II (plot (b)). Each point in $v_1/v_2$ corresponds to a different bin in $(\theta'_{c.m.}, \varphi'_{c.m.})$. The full circles correspond to out-of-plane data. The insert in plots (a) and (b) is a zoom on the in-plane data (triangles), all concentrated at small values of $v_1/v_2$. The straight line refers to the fit performed on all data points (in-plane + out-of-plane).
}
\label{fig-fitsf}
\end{figure}

Then a linear fit of the $\Psi_0$ points, based on eq.(\ref{eq-let2}), yields the two structure functions $\plltt$ and $\plt$. At fixed $q_{c.m.}$ and $\epsilon$, the coefficients $v_1$ and $v_2$ depend only on the Compton angles  $(\theta'_{c.m.}, \varphi'_{c.m.})$. The good lever arm in $v_1$ and $v_2$ is provided by the large coverage in these two angles.  Figure~\ref{fig-fitsf} represents the fit in terms of  $\Psi_0/v_2$ versus the ratio $v_1/v_2$. $\plt$ is given by the intercept and $\plltt$ by the slope of the straight line fit. The rather good $\chi^2$ (cf. Table~\ref{tab-sf-lex}) confirms that higher-order terms  $O(q'^2_{c.m.})$ are small below the pion threshold. The values obtained for the two structure functions with their errors are reported in Table~\ref{tab-sf-lex}.

\begin{table}[htb]
\caption{\label{tab-sf-lex} The structure functions obtained by the LEX analysis. The first error is statistical, the second one is the total systematic error. The reduced $\chi^2$ of the fit and the number of degrees of freedom are also given.
 }
\begin{ruledtabular}
\begin{tabular}{ccccc}
\multicolumn{5}{c}{ This experiment, LEX Analyses}  \\ 
\hline
data & $Q^2$  & $\epsilon$  &
$\plltt$ & $\plt$ \\
set & (GeV$^2$)  & \ & (GeV$^{-2}$) & (GeV$^{-2}$) \\
\hline
I-a  & 0.92 & 0.95 & 1.77 $\pm \, 0.24 \pm 0.70$ & -0.56 $\pm \, 0.12 \pm 0.17$ \\
II   & 1.76 & 0.88 & 0.54 $\pm \, 0.09 \pm 0.20$ & -0.04 $\pm \, 0.05 \pm 0.06$ \\
\hline
I-a & \multicolumn{4}{c}{$\chi^2_{min}$ = 1.22 for 32 d.o.f.}  \\ 
II  & \multicolumn{4}{c}{$\chi^2_{min}$ = 1.50 for 31 d.o.f.}  \\ 
\end{tabular}
\end{ruledtabular}
\end{table}

The statistical errors are provided by the $\chi^2$ minimization. The fit can be performed on out-of-plane and in-plane data separately: the two corresponding types of results agree within statistical errors for data set I-a, but only within total errors (statistical + systematic) for data set II.

For the systematic errors on the structure functions, one proceeds as for the cross section. The same sources of uncertainty are considered: 1) overall absolute normalization, 2) beam energy, 3) horizontal angles of the detected particles. The only difference is that the normalization error is now enlarged to $\pm 3$\% to account for uncertainties in the proton form factors, as explained in section~\ref{subsec-ff}. The LEX analysis is redone using several sets of modified cross section data, as described in section~\ref{subsec-systerr-cs}. The deviations of the structure functions w.r.t. the nominal analysis are recorded for all these cases, and finally added in quadrature. Detailed contributions to the systematic error  are given in Table~\ref{tab-sf-err1} of Appendix~\ref{app-2}. \newline
A number of complementary systematic checks were performed, e.g. by changing the analysis cuts, or the phase-space points for the cross section, etc. The physics results obtained in these studies all stay within  the systematic error bars of Table \ref{tab-sf-lex}.

\subsubsection{DR analysis} \label{subsec-dr-ana}

This analysis is based on the DR formalism introduced in section~\ref{subsec-dr-theo}. It is applied to the three data sets I-a, I-b and II separately. 
%
%
%
%
%
The restriction to stay below the pion threshold is now removed. Strictly speaking, the DR formalism provides a rigorous treatment of the VCS amplitude only up to the two-pion threshold ($W$=1.21 GeV). However, the two-pion contribution is still small just above threshold; the upper limit in $W$ in our analysis is taken at  $W$=1.28 GeV, considering that the model calculation is able to describe the experimental data in this energy range (see, for example, Fig.3 of ref.~\cite{Laveissiere:2008zn}).
%
 The cross-section data of Tables~\ref{tab-cs-1} to \ref{tab-cs-5} are included. Globally, two different domains  in $W$ are involved:  1) the region of the $\Delta (1232)$ resonance for data set I-b,  2) the region essentially below the pion threshold, with a small extension above, for data sets I-a and II (cf.  Fig.~\ref{fig-rangeinw}).

\begin{figure}[htb]
\includegraphics[width=8.3cm]{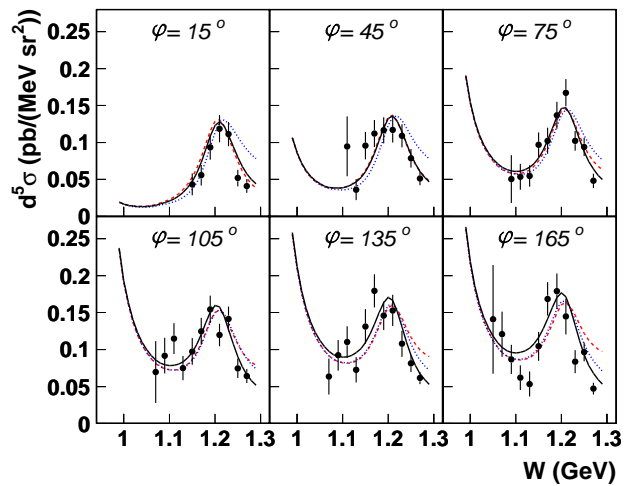}
\caption{(Color online) Data set I-b. The $(ep \to ep \gamma )$ cross section at fixed $\cos \theta_{c.m.}=-0.975$ and six bins in $\varphi$ (from Table \ref{tab-cs-5}). The curves show the DR model calculation for parameter values  $( \lama, \lamb )$=(0.70, 0.63) GeV (solid), (1.20, 0.63) GeV  (dashed), (0.70, 1.00) GeV  (dotted).
 }
\label{fig-secdelta}
\end{figure}

Some of our cross-section measurements above the pion threshold are depicted in Figs. \ref{fig-cs-above} and \ref{fig-secdelta}.  Figure \ref{fig-secdelta} clearly shows the excitation of the $\Delta(1232)$ resonance. The DR model gives a good description of the data and the various curves illustrate the sensitivity to the model parameters. The  peak position is located at a lower mass  than the $\Delta$ mass, a feature due to the VCS-BH interference.  Figure \ref{fig-deltam-lex} shows how the DR model reproduces the $q'_{c.m.}$-dependence of the quantity  $\Delta {\cal M}$ introduced in section \ref{subsec-lex-ana}: the flat behavior below the pion threshold is followed by a rise near the $\Delta$ resonance, where the higher-order terms become dominant.

The DR formalism incorporates the BH+Born cross section, and a Non-Born part which contains the free parameters $\lama$ and $\lamb$ (cf. section \ref{subsec-dr-theo}). The analysis method consists in fitting these two parameters by a $\chi^2$ minimization (called the ``DR fit'') which compares the model cross section to the measured one. The minimization cannot be solved analytically;  the $\chi^2$ is computed on the nodes of a grid in ($\lama, \lamb)$ and its minimum is found numerically. For each data set a clear and single minimum is found, with a  reasonable  $\chi^2_{min}$ value.  The fitted values for $( \lama, \lamb )$  corresponding to our three independent data sets are reported in Table \ref{tab-dr-lalb}.

The statistical errors on ($\lama, \lamb )$ are given by the standard error ellipse at ($\chi^2_{min}+1$). The systematic errors are treated exactly as in the LEX method, i.e. by finding the solution in ($\lama, \lamb)$  for modified  cross-section sets, and summing quadratically the resulting variations w.r.t. the nominal analysis. Table \ref{tab-lalb-errsys} of Appendix~\ref{app-2} gives these detailed contributions. One will note that the systematic errors appear to be much smaller in the case of data set I-b. This may come from the different phase space coverage of this data set (in $\theta_{c.m.}$ and W), inducing a different sensitivity to the sources of errors. Remarkably, all our fitted values of the  ($\lama, \lamb )$ parameters lie in a  narrow range: [0.63,0.79] GeV, indicating that the asymptotic part of the GPs ($\Delta \alpha,\Delta \beta$ of eqs.(\ref{eq-dr-beta-0}) and (\ref{eq-dr-alpha-0}))  behaves roughly as a single dipole in the $Q^2$-range of 1-2 GeV$^2$.

\begin{table}[htb]
\caption{\label{tab-dr-lalb} 
The fitted dipole mass parameters $\lama$ and $\lamb$ for the three independent data sets. The first and second errors are statistical and total systematic, respectively. The reduced $\chi^2$ of the fit and the number of degrees of freedom are also given.
}
\begin{ruledtabular}
\begin{tabular}{ccc}
 data & $\lama$  & $\lamb$  \\
 set &  (GeV) & (GeV) \\
\hline
I-a & 0.741 $\pm \, 0.040 \pm 0.175$ & 0.788  $\pm \, 0.041 \pm 0.114$  \\ 
I-b & 0.702 $\pm \, 0.035 \pm 0.037$ & 0.632  $\pm \, 0.036 \pm 0.023$  \\
II &  0.774 $\pm \, 0.050 \pm 0.149$ & 0.698  $\pm \, 0.042 \pm 0.077$  \\
\hline
I-a & \multicolumn{2}{c}{$\chi^2_{min}$ = 1.49 for 164 d.o.f.}  \\ 
I-b & \multicolumn{2}{c}{$\chi^2_{min}$ = 1.34 for 328 d.o.f.}  \\ 
II  & \multicolumn{2}{c}{$\chi^2_{min}$ = 1.31 for 151 d.o.f.}  \\ 
\end{tabular}
\end{ruledtabular}
\end{table}

Once we have the fitted values of $\lama$ and $\lamb$, the DR model is able to calculate the scalar GPs and the scalar part of the structure functions (defined in Eq.(\ref{eq-sf12})) at the $Q^2$ under consideration. The full structure functions $\plltt$ and $\plt$ are then formed by adding the spin part. This last is parameter-free, since all spin GPs are fixed in the DR model. The complete DR calculation is done separately for each data set using the inputs of Table~\ref{tab-dr-lalb}. The results for $\plltt$ and $\plt$  are given in Table \ref{tab-sf-dr}.  The results for $\pll$ alone are given in Table~\ref{tab-dr-other}, where we have also reported the DR value of the spin part $\ptt$.

\begin{table}[htb]
\caption{\label{tab-sf-dr} 
The VCS structure functions obtained by the DR analysis. The first error is statistical, the second one is the total systematic error. 
 }
\begin{ruledtabular}
\begin{tabular}{ccccc}
\multicolumn{5}{c}{ This experiment, DR Analysis}  \\ 
\hline
data & $Q^2$  & $\epsilon$  &
$\plltt$ & $\plt$ \\
set & (GeV$^2$)  & \ & (GeV$^{-2}$) & (GeV$^{-2}$) \\
\hline
I-a  & 0.92 & 0.95 & 1.70 $\pm \, 0.21 \pm 0.89$ & -0.36 $\pm \, 0.10 \pm 0.27$ \\
I-b  & 0.92 & 0.95 & 1.50 $\pm \, 0.18 \pm 0.19$ & -0.71 $\pm \, 0.07 \pm 0.05$ \\
II   & 1.76 & 0.88 & 0.40 $\pm \, 0.05 \pm 0.16$ & -0.09 $\pm \, 0.02 \pm 0.03$ \\
\end{tabular}
\end{ruledtabular}
\end{table}

\begin{table}[htb]
\caption{\label{tab-dr-other} Our result for the $P_{LL}$ structure function from the DR analysis. The value of $P_{TT}$ is the parameter-free DR prediction. 
 }
\begin{ruledtabular}
\begin{tabular}{cccc}
data & $Q^2$  & $P_{LL}$ (exp.) & $P_{TT}$ (theory) \\
set & (GeV$^2$)  & (GeV$^{-2}$) & (GeV$^{-2}$) \\
\hline
I-a  & 0.92 & 1.19  $\pm 0.21$   $\pm 0.89$  &  -0.485  \\
I-b  & 0.92 & 0.99  $\pm 0.18$   $\pm 0.19$  &  -0.485  \\
II   & 1.76 & 0.24  $\pm 0.05$   $\pm 0.16$  &  -0.142  \\
\end{tabular}
\end{ruledtabular}
\end{table}

To obtain the statistical and systematic errors in Tables \ref{tab-sf-dr} and \ref{tab-dr-other}, the errors on ($\lama, \lamb )$ are propagated to the structure functions, using the model calculation. One will note that the DR model exists in several versions, each one using a different set of MAID multipoles for pion electroproduction. Our analyses were done using MAID 2000. With more recent multipole sets (MAID 2003 or 2007), the ($\epg$) cross section in the DR model changes by $\sim$ 1-2\% in our kinematics. Therefore the results presented here are not expected to change noticeably with the version update; they should stay largely within the quoted statistical error.

\subsubsection{Consistency between the different analyses}
 \label{subsec-discuss-sf}

\begin{figure}[htb]
\includegraphics[width=8.3cm]{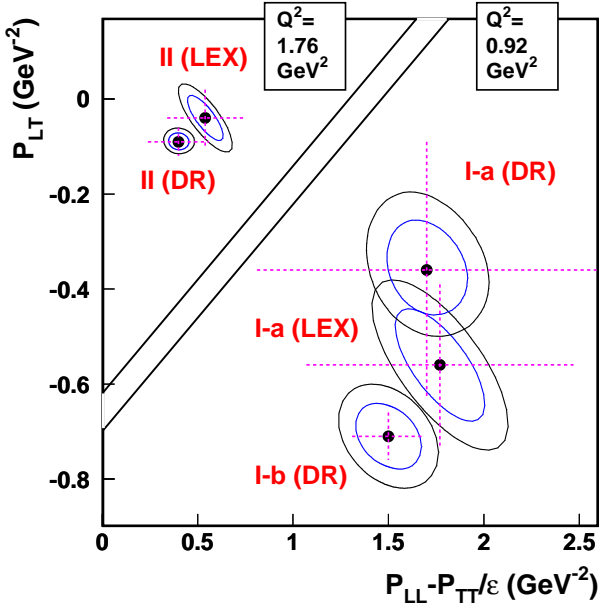}
\caption{(Color online) The structure functions obtained by the two methods (LEX and DR) at $Q^2=0.92$ GeV$^2$ and 1.76 GeV$^2$. Each $Q^2$ corresponds to a solid-line polygon as marked. For each point, the inner ellipse is the contour at ($\chi^2_{min}+1$), yielding the standard deviation  on each structure function independently.  The outer ellipse at  ($\chi^2_{min}+2.3$)  corresponds to a probability of 68\% that both structure functions are inside the contour simultaneously.  The statistical errors quoted in Tables \ref{tab-sf-lex} and \ref{tab-sf-dr} are given by the boundaries of the inner contour. Dotted crosses give the size of the systematic error.
}
\label{fig-sf-consistency}
\end{figure}

The consistency between the two types of analysis, LEX and DR, can be tested only on $\plltt$ and $\plt$ (not on the GPs themselves), because these two structure functions  are the only direct outcome of the LEX analysis. In  Fig.~\ref{fig-sf-consistency} we give a comprehensive view of all our measurements of the structure functions, for the three independent data sets and the two analysis methods. The representation in the form of standard error ellipses indicates that error correlations between the two structure functions are larger in the LEX case than in the DR case.

At $Q^2=0.92$ GeV$^2$, the three  measurements of $\plltt$ agree very well. The agreement is less good on the three values of $\plt$, in particular between  the separate DR extractions of $\plt$ from data sets I-a and I-b, i.e. essentially below and above the pion threshold. These values become however compatible within total errors, including systematics. As a side remark, we note that a single DR analysis on the whole $W$-range would be possible, by joining together data sets I-a and I-b, but it would mask these different results for  $\plt$. At $Q^2=1.76$ GeV$^2$, the LEX and DR results are in mild agreement.

Overall, Fig.~\ref{fig-sf-consistency} shows a rather good consistency between the two types of extraction methods, LEX and DR, at each $Q^2$. We also point out that the systematic error generally dominates in our physics results ( data sets I-a and II). This feature can be understood already at the cross section level, where the size of the systematic error ($\sim$ 7\% of the cross section) is about half the size of the expected GP effect (10-15\% of the cross section below the pion threshold). 

At one given $Q^2$, our LEX and DR results are obtained in most cases from non-independent, partially overlapping data sets. Therefore we do not propose any averaging of the points shown in  Fig. \ref{fig-sf-consistency}, at each $Q^2$. Only the (I-b(DR)) and (I-a(DR)) results are truly independent and could possibly be averaged.


\subsubsection{$Q^2$-dependence of the structure functions}
 \label{subsec-discuss-sf-q2dep}

Most of the theoretical models for GPs (section \ref{subsec-theo-models}) have a validity domain limited to low energies and low $Q^2$.  At $Q^2$ values of 1-2 GeV$^2$, the only relevant confrontation of experimental data is with the dispersive approach, so we will focus on this model. ChPT will still be included as a reference in the lower-$Q^2$ region.

\begin{figure}[htb]
\includegraphics[width=8.3cm]{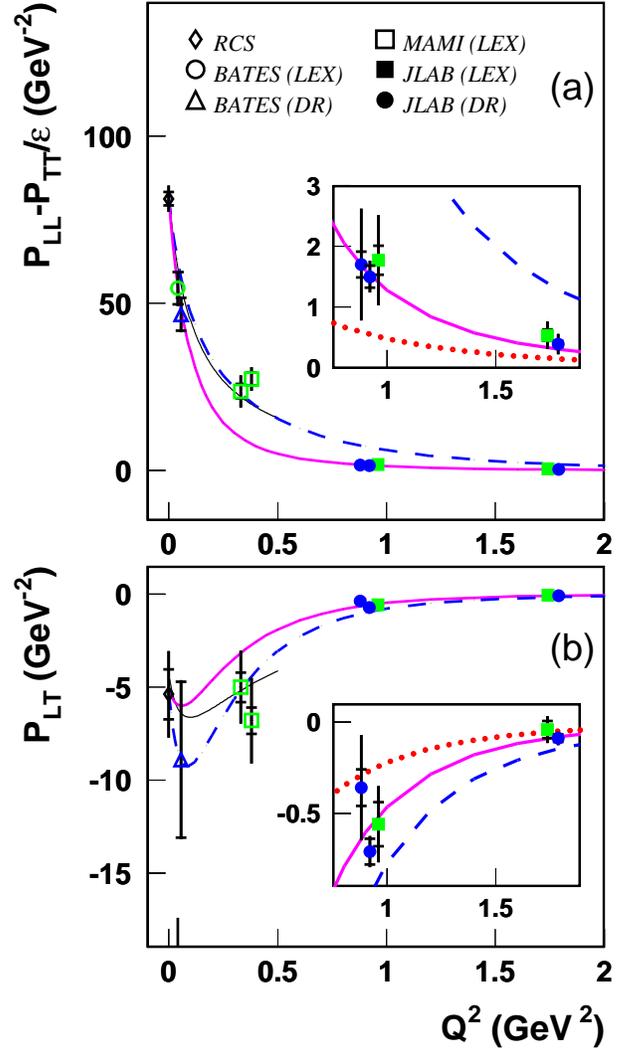}
\caption{(Color online) The structure functions $\plltt$ (a) and $\plt$ (b) measured at Bates \cite{Bourgeois:2011te}, MAMI \cite{Roche:2000ng,Janssens:2008qe} and JLab \cite{Laveissiere:2004nf} (this experiment). The RCS point deduced from \cite{OlmosdeLeon:2001zn} is also included. The inserts are a zoom in the $Q^2$-region of this experiment. Some points are slightly shifted in abscissa  for visibility.  The inner (outer) error bar on the points is statistical (total). The thin solid curve is the HBChPT calculation \cite{Hemmert:1999pz}. The other curves show DR calculations \cite{Pasquini:2001yy} performed with various sets of parameters. Plot (a): $\lama=0.7$ GeV (thick solid),  $\lama=1.79$ GeV (dashed). Plot (b): $\lamb =0.7$ GeV (thick solid),  $\lamb =0.5$ GeV (dashed). $\epsilon=0.9$ is chosen to draw the curves  for $\plltt$. The dotted curve in the inserts is the spin part as given by the DR model: $-P_{TT}/0.9$ (upper plot) and $P_{LTspin}$ (bottom plot).
}
\label{fig-sf1}
\end{figure}

 Figure~\ref{fig-sf1} shows the structure functions obtained in this experiment, together with the other existing measurements, and model calculations. The main strength of the JLab data is to have enlarged considerably the measured  $Q^2$-range, allowing to put in perspective different regions of four-momentum transfer, ``high'' and ``low''.


The experimental data follow the global trend of the models, i.e. a more or less continuous fall-off for $\plltt$, and for $\plt$ a rather flat behavior in the low-$Q^2$ region followed by an asymptotic trend to zero. At low $Q^2$ the data are in good agreement with HBChPT at $O(p^3)$~\cite{Hemmert:1999pz} (thin solid curve). The DR model does not give a parameter-free prediction of $\plltt$ and $\plt$. To draw the DR curves in  Fig.~\ref{fig-sf1} we have fixed the dipole mass parameters $\lama$ and $\lamb$ of eqs.(\ref{eq-dr-beta-0}) and (\ref{eq-dr-alpha-0}), and further assumed that they are constant versus $Q^2$. This is a simplification, only aiming at a simple graphical representation; as explained in section \ref{subsec-dr-theo}, the DR model has no such constraint intrinsically. The solid curve shows the DR calculation for typical parameter values obtained in our experiment: $\lama = \lamb = 0.7$ GeV. The dashed curve shows the DR calculation for other parameter values, which agree better with some of the low-$Q^2$ data.

\begin{figure}[htb]
\includegraphics[width=8.3cm]{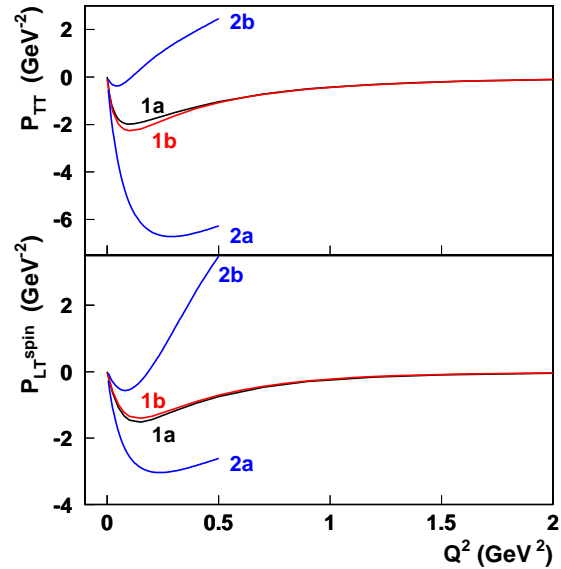}
\caption{(Color online) Theoretical predictions for the spin part of the measured structure functions. 
The DR curves labelled 1a and 1b are calculated with the MAID 2000 and MAID 2003 multipoles, respectively. The HBChPT curves 2a and 2b are obtained at $O(p^3)$ \cite{Hemmert:1999pz} and $O(p^4)$ \cite{Kao:2002cn}, respectively. They are drawn  up to an arbitrary value of $Q^2_{max}=0.5$ GeV$^2$.
}
\label{fig-sf2}
\end{figure}

As a general statement, there is no such single DR curve which goes well through all the data points, over the whole $Q^2$-range. It means that a single dipole function for the unconstrained parts $\Delta \alpha$ and $\Delta \beta$ of eqs.(\ref{eq-dr-beta-0}) and (\ref{eq-dr-alpha-0}) is too limiting. This is especially true for the first structure function $\plltt$: all measurements are compatible with the thick solid curve ($\lama = 0.70$ GeV), except near $Q^2 \simeq 0.3$ GeV$^2$ (MAMI points) where an enhancement is observed in the data. This feature becomes more pronounced when dealing with the electric GP, and will be further discussed in the next section. It should be noted that all measurements are performed at high $\epsilon$, around 0.9, except the MAMI points which are at $\epsilon \simeq 0.6$. However this change in  $\epsilon$ can hardly account for the observed enhancement near $Q^2 \simeq 0.3$ GeV$^2$, since $\ptt$ is expected to be a very small quantity (cf.  Fig.~\ref{fig-sf2}).


For the second structure function $\plt$, we note that at our highest $Q^2$ (1.76 GeV$^2$) the measured value is almost zero, within errors (especially the LEX result). It is suggestive of the limitations of the present extraction methods, w.r.t. higher momentum transfers. Turning back to the low-$Q^2$ region,  most models predict an extremum of $\plt$. This feature is more or less confirmed by experiment, but error bars are still large and can hopefully be reduced in the future. The global behavior of $\plt$ essentially reflects the $Q^2$-dependence of the magnetic GP $\bem$, which will be discussed in the next section.

There are no measurements of the spin part of the structure functions, $\ptt$ and $P_{LTspin}$. Theoretical estimates are given in  Fig.~\ref{fig-sf2}. In HBChPT the spin GPs have been calculated up to order $O(p^4)$ \cite{Kao:2002cn,Kao:2004us}, i.e. one more order than for the scalar GPs, but the calculation does not show good convergence (cf. curves 2a and 2b in  Fig.~\ref{fig-sf2}). This should be kept in mind when considering  Fig.~\ref{fig-sf1}: the good agreement between the low-$Q^2$ data and HBChPT  at $O(p^3)$ may be accidental and not so well verified at next order. In  Fig.~\ref{fig-sf2} the DR calculation gives (parameter-free) spin structure functions of very small magnitude, decreasing rapidly with $Q^2$ and almost independent of the set of $\pi N$ multipoles. This contribution is also drawn in  Fig.~\ref{fig-sf1} as the dotted lines. It amounts to 35-50\% of the measured structure functions in the $Q^2$-region of our experiment, but this percentage is much smaller at lower $Q^2$. Experimental information on the spin GPs would be very valuable, but very little is available, due to the difficulty of such experiments \cite{Merkel:2000,Janssens:2007th,Doria:2007th}.

\subsection{The electric and magnetic GPs} \label{subsec-discuss-gp}

The data of this experiment allow the extraction of the electric and magnetic GPs of the proton at $Q^2$ = 0.92 and 1.76 GeV$^2$, as an ultimate step of our analyses. In the DR formalism, these GPs are calculated in a straightforward way, once the ($\lama, \lamb$) parameters are known (``direct DR extraction''). On the other hand, in the LEX formalism there is no such direct determination of the GPs. The spin structure functions $\ptt$ and $P_{LTspin}$ have first to be subtracted from the measured ones, $\plltt$ and $\plt$, using  a model. For this task it is most natural to choose the DR model, especially in our $Q^2$-range. Once this subtraction is done, the last step is to remove the $Q^2$-dependence due to the electric form factor $G_E^p$ in the scalar part (cf.eq.(\ref{eq-sf12})). Our results for $\ale(Q^2)$ and $\bem(Q^2)$, following this procedure, are reported in Table \ref{resultsgp}, together with the results of the direct DR extraction.

\begin{table}[htb]
\caption{\label{resultsgp} 
The  electric and magnetic GPs extracted in this experiment
at $Q^2$= 0.92 GeV$^2$  and 1.76 GeV$^2$.
The first error is statistical. The second one is the total systematic error,  obtained by propagating the errors on ($\lama , \lamb$) of Table \ref{tab-sf-dr}  (for the DR analysis) or the errors on the structure functions  of Table \ref{tab-sf-lex} (for the LEX analysis).
}
\begin{ruledtabular}
\begin{tabular}{cccc}
data   & $Q^2$ & $\alpha_E(Q^2)$  
& $\beta_M(Q^2)$  \\
set & (GeV$^2$) & ($10^{-4}$ fm$^3$) &  ($10^{-4}$ fm$^3$) \\
 \hline
\multicolumn{4}{c}{DR analysis} \\
 \hline
I-a     & 0.92    & 1.02  $\pm \, 0.18 \pm 0.77$    
                  & 0.13  $\pm \, 0.15 \pm 0.42$  \\ 
I-b     & 0.92    & 0.85  $\pm \, 0.15 \pm 0.16$ 
                  & 0.66  $\pm \, 0.11 \pm 0.07$  \\
II      & 1.76    & 0.52  $\pm \, 0.12 \pm 0.35$
                  & 0.10  $\pm \, 0.07 \pm 0.12$   \\
 \hline
\multicolumn{4}{c}{LEX analysis + [spin part subtraction by DR]} \\
 \hline
I-a     & 0.92    & 1.09  $\pm \, 0.21 \pm 0.60$    
                  & 0.42  $\pm \, 0.18 \pm 0.26$  \\ 
II      & 1.76    &  0.82   $\pm \, 0.20 \pm 0.44$
                  & -0.06   $\pm \, 0.17 \pm 0.20$   \\
\end{tabular}
\end{ruledtabular}
\end{table}

\begin{figure}[htb]
\includegraphics[width=8.3cm]{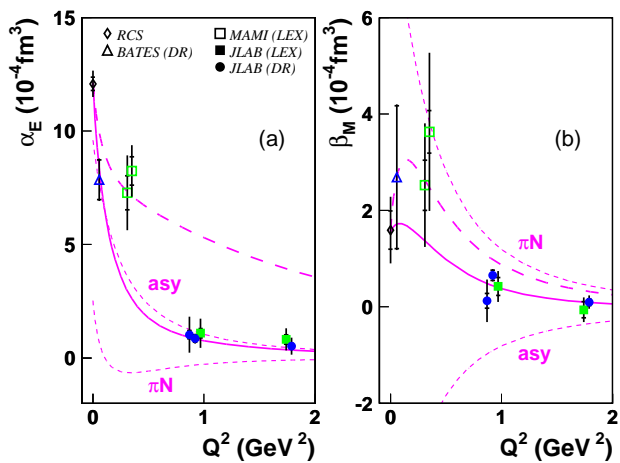}
\caption{(Color online) The world data on the electric GP (a) and the magnetic GP (b), with statistical (inner) and total (outer) error bar. The solid curve is the DR calculation drawn for typical parameter values obtained in our experiment: $\lama = 0.70$ GeV (left) and $\lamb = 0.70$ GeV (right). The short-dashed curves show the two separate contributions to this calculation: the pion-nucleon intermediate states (curve labelled ``$\pi N$'') and the [asymptotic + beyond $\pi N$] contribution (curve labelled ``asy''). The long-dashed curve  is the full DR calculation for other parameter values: $\lama = 1.79$ GeV (left) and  $\lamb = 0.51$ GeV (right).  
}
\label{fig-ab1}
\end{figure}

 Figure~\ref{fig-ab1} summarizes the existing measurements of $\ale (Q^2)$ and $\bem (Q^2)$ of the proton. It is clear that this picture is to some extent (DR)model-dependent; for consistency, theoretical curves are drawn only for this particular model. Similarly to  Fig.~\ref{fig-sf1}, a single DR curve cannot reproduce all the experimental data, and an enhancement can be seen in the region of the MAMI points. In a recent paper interpreting the GPs in the light-front formalism~\cite{Gorchtein:2009qq}, the electric GP is described by adding to the DR calculation a Gaussian contribution centered near $Q^2$=0.3 GeV$^2$. With this parametrization, one is able to reproduce the  measured $\plltt$ over the full $Q^2$-range, and the induced electric polarization in the nucleon is shown to extend to larger transverse distances. However in \cite{Gorchtein:2009qq} no clear physical origin is associated to this additional and intriguing structure in $\ale(Q^2)$.

In the DR model, each scalar GP is the sum of two terms (cf. section~\ref{subsec-dr-theo}): the dispersion integrals, saturated by the $\pi N$ contribution, and the [asymptotic + beyond $\pi N$] contribution $\Delta \alpha$ or $\Delta \beta$.  Figure~\ref{fig-ab1} shows these two contributions separately (short-dashed curves). For the electric GP the [asymptotic + beyond $\pi N$] term is by far dominant at every $Q^2$. For the magnetic GP the two contributions are large and of opposite sign. The $\pi N$ dispersive integral is of paramagnetic nature, namely via the formation of the $\Delta$(1232) resonance. The diamagnetism (``asy'' curve) arises from the $\Delta \beta$ term associated with the exchange of the $\sigma$-meson (=$[ \pi\pi ]_0$) in the $t$-channel. The two terms strongly cancel, leading to an overall small polarizability, and a more or less pronounced extremum in the low-$Q^2$ region.

 
It is well known that nucleon polarizabilities are closely linked to the mesonic cloud, an essential ingredient of nucleon structure since the first ChPT calculation of $\ale$ and $\bem$  in RCS~\cite{Bernard:1993bg}. The mesonic cloud is also expected to play an important role in the GPs. The measured value of the mean square electric polarizability radius of the proton $\langle r^2_{\alpha} \rangle $, of about 2 fm$^2$~\cite{Bourgeois:2011te}, clearly indicates that what is probed in VCS is a large-size structure. The non-trivial shape observed for the electric GP over the full $Q^2$-range calls for further understanding, and more measurements of the scalar GPs in the $Q^2$-region of [0-1] GeV$^2$ are needed to get a clearer picture. Such measurements are underway at MAMI~\cite{Merkel:2009}.


\section{Conclusion} \label{sec-concl}

The JLab E93-050 experiment was one of the first-generation VCS experiments, dedicated in part to the measurement of the generalized polarizabilities of the proton at high momentum transfer. It was a challenging task to exploit the Hall A equipment in its commissioning phase to study the exclusive process $ep \to ep \gamma$ and accurately measure its small cross section. Two methods have been used to extract the physics observables: one model-independent based on the LEX, and one model-dependent based on the DR formalism. The results of the two methods show good consistency at the level of the VCS structure functions $\plltt$ and $\plt$. 

The data obtained in this experiment allow in a unique way to put in perspective the regions of high and low momentum transfer. The results are an essential piece to build a more complete picture of the electric and magnetic GPs of the proton as a function of $Q^2$, i.e. ultimately the nucleon's polarization response as a function of the distance scale. The electric GP does not seem to have a smooth fall-off, and the behavior of the magnetic GP quantifies the detailed contributions of para- and diamagnetism in the proton. Experimental data are still scarce, and more measurements are desirable in order to improve our understanding of these fundamental observables. This is especially true at low $Q^2$, where the prominent role of the mesonic cloud can be probed. New VCS experiments,  together with new RCS experiments and theoretical developments in the field, should provide a step forward in our understanding of the electromagnetic structure of the nucleon.


\ \\

We thank the JLab accelerator staff and the Hall~A technical staff for their dedication in performing this experiment and making it a success. This work was supported by DOE contract DE-AC05-84ER40150 under which the Southeastern Universities Research Association (SURA) operates the Thomas Jefferson National Accelerator Facility. We acknowledge additional grants from the US DOE and NSF, the French CNRS and CEA,  the Conseil R\'egional d'Auvergne, the FWO-Flanders (Belgium) and the BOF-Ghent University. We thank the INT (Seattle) and ECT* (Trento) for the organization of VCS workshops.

\bibliography{ref444}


\appendix
\section{Kinematics and Binning} 
\label{app-1}


The GPs depend on $q_{c.m.}$, or equivalently on the four-momentum transfer squared $Q^2$  taken in the limit  $q'_{c.m.} \to 0$ \cite{Guichon:1998xv}. This variable is defined as ${\tilde Q}^2 = 2 \nucleonmass \cdot ( \sqrt{ \nucleonmass ^2 + { q_{c.m.}}^{ 2}} - \nucleonmass )$. It has been denoted $Q^2$ throughout the paper for simplicity.

\begin{figure}[htb]
\begin{center}
\includegraphics[width=6.0cm]{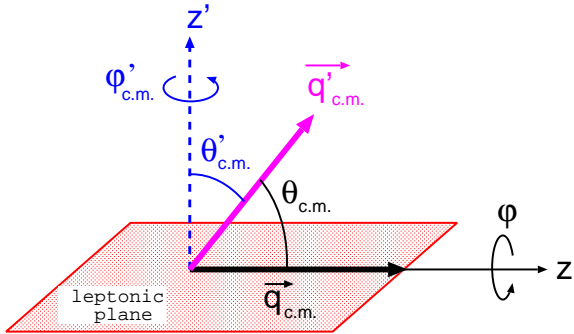}  
\end{center}
\caption{(Color online) The angular systems to measure the polar and azimuthal angles of the momentum vector ${\vec q} \, '_{c.m.}$ in the ($\gamma p$) center-of-mass.
}
\label{fig-axisrot}
\end{figure}

Two angular systems in the \CM are described in section~\ref{subsec-cross1}: ``standard'' and ``rotated''. In the standard system, the polar angle $\theta_{c.m.}$ is measured w.r.t. the $z$ axis aligned with the  $\vec q_{c.m.}$ vector. In the rotated system, the polar angle  $\theta \, ' _{c.m.}$ is measured w.r.t. the  $z'$ axis orthogonal to the leptonic plane; see  Fig.~\ref{fig-axisrot}.  In-plane kinematics correspond to $\theta \, ' _{c.m.}=90^{\circ}$, or to $\varphi=0^{\circ}$ and $180^{\circ}$. The conversion formulae between  the two systems are the following: 
\begin{eqnarray}
\begin{array}{lll}
\cos  \theta \, ' _{c.m.} & = &  \sin \theta_{c.m.} \cdot \sin \varphi \ , \\
\cos \varphi \,' _{c.m.} & = &  \cos \theta_{c.m.} \  /  \  \sin \theta \, ' _{c.m.}  \ , \\
\sin \varphi \, ' _{c.m.} & = &  \sin \theta_{c.m.} \cdot \cos \varphi \ /  \  \sin \theta \, ' _{c.m.}  \ . \\
\label{eq-rota}
\end{array}
\end{eqnarray} 

Table \ref{tab-bins} gives the binning information for the three data sets. The variables which are kept fixed are different for (I-a and II) and for (I-b). The three first bins in $q'_{c.m.}$ are below the pion threshold, the other ones are above. For $\theta'_{c.m.}$, the up-down symmetry w.r.t. the leptonic plane allows to sum the intervals as indicated ($\cup$ symbol). This up$+$down sum is also applied in the $\varphi$-bins of data set I-b.

\begin{table}[htb]
\caption{\label{tab-bins} The phase-space variables (first column), the bin size and  the points $P_0$ at which the cross section is determined (second column). These points are chosen at the middle of the bin, except for the last bin in $q'_{c.m.}$ and the first bin in $\theta'_{c.m.}$.
 }
\begin{ruledtabular}
\begin{tabular}{ll} 
\multicolumn{2}{c}{ Data sets I-a and II }  \\
\hline
$\epsilon$ & fixed at 0.950 (0.879) for data set I-a (II) \\
$q_{c.m.}$   & fixed at 1.080 (1.625) GeV/c for data set I-a (II) \\
$q'_{c.m.}$  & 6 bins: regular from 30  to 180, + [180, 250] MeV/c \\
 \         & 6 points: 45, 75, 105, 135, 165,   215  MeV/c \\
$\theta'_{c.m.}$ & 2 bins: $[0^{\circ}, 60^{\circ}] \cup [120^{\circ},180^{\circ}]$; $[60^{\circ}, 90^{\circ}] \cup [90^{\circ},120^{\circ}]$  \\
 \             & 2 points: 40$^{\circ}$ (=out-of-plane) and  90$^{\circ}$ (=in-plane) \\  
$\varphi'_{c.m.}$ & 20 bins: regular  from -180$^{\circ}$ to +180$^{\circ}$ \\ 
 \              & 20 points:  $-171^{\circ}, -153^{\circ}, -135^{\circ}$, ...  , $ +153^{\circ}, +171^{\circ}$ \\
\hline
\multicolumn{2}{c}{ Data set I-b \ \ (cf. ref.~\cite{Laveissiere:2008zn}) }  \\
\hline
$k_{lab}$ & fixed at 4.032 GeV  \\
$Q^2$     & fixed at 1.0 GeV$^2$ \\
$W$ & 15 bins: regular  from 0.98 GeV to 1.28 GeV  \\
 \  & 15 points:  $0.99, 1.01, 103$ , ... ,  $1.25, 1.27$ GeV  \\
$\cos \theta_{c.m.}$ & 3 bins: [-1, -0.95], [-0.95, -0.80], [-0.80, -0.50]  \\ 
 \                 & 3 points: -0.975, -0.875, -0.650  \\ 
$\varphi$ & 6 bins:  regular from 0 to 180$^{\circ}$  \\ 
 \        & 6 points: 15$^{\circ}$, 45$^{\circ}$, 75$^{\circ}$, 105$^{\circ}$, 135$^{\circ}$, 165$^{\circ}$  \\
\end{tabular}
\end{ruledtabular}
\end{table}

\section{Systematic errors} 
\label{app-2}

This Appendix gives details on the systematic errors on the primary observables  extracted from the cross section data: the structure functions in the LEX analysis (Table \ref{tab-sf-err1}) and the  ($\lama, \lamb$) parameters in the DR analysis (Table \ref{tab-lalb-errsys}). The systematic errors on all other extracted observables are derived from these ones by error propagation.
The origin of the 3\% normalization error in these Tables is explained in section \ref{subsec-ff}.

\begin{table}[ht]
\caption{\label{tab-sf-err1} Detailed systematic errors on the structure functions in the LEX analysis. To obtain the total systematic error, the three partial contributions are first symmetrized ($+/-$) and then added in quadrature.
 }
\begin{ruledtabular}
\begin{tabular}{lcc}
source of &  $\Delta ( \plltt )$  &  $\Delta ( P_{LT} ) $  \\
systematic error &
 (GeV$^{-2}$) &
 (GeV$^{-2}$)  \\
\hline
\multicolumn{3}{c}{Data Set I-a}  \\ 
 \hline
normalization ($\pm$ 3 \%) & +0.505 -0.505 & +0.046 -0.046  \\
beam  energy ($\pm$ 2 MeV) & +0.391 -0.354  & +0.132 - 0.024 \\
spec.angle ($\pm$ 0.5 mr) &+0.301 -0.301  & +0.148 -0.148 \\
Total syst. error &  $\pm$ 0.696  &  $\pm$ 0.174  \\
\hline
\multicolumn{3}{c}{Data Set II}  \\ 
\hline
normalization ($\pm$ 3 \%) & +0.142 -0.142  & +0.004 -0.004 \\
beam  energy ($\pm$ 2 MeV) & +0.139 -0.074  & +0.017 -0.005 \\
spec.angle ($\pm$ 0.5 mr) & +0.096 -0.096  & +0.054 -0.054 \\
Total syst. error &  $\pm$ 0.202  & $\pm$ 0.055 \\
\end{tabular}
\end{ruledtabular}
\end{table}

\begin{table}[htb]
\caption{\label{tab-lalb-errsys} Detailed systematic errors on the parameters $\lama$ and $\lamb$ in the DR analysis. To obtain the total systematic error, the three partial contributions are first symmetrized ($+/-$) and then added in quadrature.
 }
\begin{ruledtabular}
\begin{tabular}{lcc}

source of &  $\Delta ( \lama )$  &  $\Delta (  \lamb  ) $  \\
system.error &
 (GeV) &
 (GeV)  \\
\hline
\multicolumn{3}{c}{Data Set I-a}  \\ 
\hline
normalization ($\pm$ 3 \%) &  +0.105 -0.099 & +0.068 -0.053   \\
beam  energy ($\pm$ 2 MeV) &  +0.179 -0.067  & +0.068 - 0.018\\
spec.angle ($\pm$ 0.5 mr) & +0.072 -0.072  & +0.087 -0.087 \\
Total syst. error &  $\pm$ 0.175  & $\pm$ 0.114   \\
\hline
\multicolumn{3}{c}{Data Set II}  \\ 
\hline
normalization ($\pm$ 3 \%) & +0.096 -0.127  & +0.057 -0.056 \\
beam  energy ($\pm$ 2 MeV) &  +0.118 -0.047  & +0.044 -0.017 \\
spec.angle ($\pm$ 0.5 mr) &  +0.052 -0.052  & +0.041 -0.041  \\
Total syst. error &  $\pm$   $\pm$ 0.149  & $\pm$ 0.077 \\
\hline
\multicolumn{3}{c}{Data Set I-b}  \\ 
\hline
normalization ($\pm$ 3 \%) & +0.031 -0.041  & +0.018 -0.014 \\
beam  energy ($\pm$ 2 MeV) &  +0.017 -0.003  & +0.011 -0.012 \\
spec.angle ($\pm$ 0.5 mr) &  +0.005 -0.005  & +0.012 -0.012  \\
Total syst. error &  $\pm$   $\pm$ 0.037  & $\pm$ 0.023 \\
\end{tabular}
\end{ruledtabular}
\end{table}

\section{Cross section Tables} 
\label{app-3}

This Appendix gives our measured values of the photon electroproduction cross section. Table ~\ref{tab-cs-1} (resp. \ref{tab-cs-2}) is for data set I-a (resp. II) below the pion threshold. Table ~\ref{tab-cs-3} (resp. \ref{tab-cs-4}) is for  data set I-a (resp. II) above the pion threshold.  In these four Tables, the errors are statistical only (in r.m.s); the systematic error is discussed in the text (section \ref{subsec-systerr-cs}).

The cross-section values for data set I-b can be found partly in ref.~\cite{Laveissiere:2008zn}, but there some settings of data set I-a were also included. Thus, for the sake of completeness, the cross-section values corresponding to the DR analysis (I-b) of the present paper are reported here in Table \ref{tab-cs-5}. It should be noted that this cross section has been determined at $Q^2=1.0$ GeV$^2$ (instead of 0.92 GeV$^2$), as part of a wider study program at $Q^2=1$ GeV$^2$ \cite{Laveissiere:2008zn}. To obtain the VCS observables at $Q^2=0.92$ GeV$^2$ from the ``I-b'' DR fit, we have explicitely assumed that the ($\lama,\lamb$) parameters have no significant variation {\it locally} in $Q^2$. Then we simply use the parameter values of Table \ref{tab-dr-lalb} (I-b part), fitted at $Q^2=1.0$ GeV$^2$, as inputs to the DR calculation of the VCS observables  at $Q^2=0.92$ GeV$^2$.

Ascii files of Tables ~\ref{tab-cs-1} to ~\ref{tab-cs-5} will be added as auxiliary files to this paper. They are also available at URLs: 
http://userweb.jlab.org/$\sim$helene/paper{\_}vcs{\_}gps{\_}2012/all-ascii-tables, and http://clrwww.in2p3.fr/sondem/E93050-tables-GPS, or upon request to the authors.



\begin{table*}[h]
\caption{ The measured $(ep \to ep \gamma)$ cross section $d^5 \sigma / d k'_{e lab} d \Omega'_{e lab} d\Omega_{\gamma c.m.}$, in pb/(GeV sr$^2$), for data set I-a below the pion threshold. The out-of-plane (resp. in-plane) data correspond to $\theta'_{c.m.}=40^{\circ}$ (resp. $90^{\circ}$). The error $\Delta \sigma$ is statistical only. The  $(ep \to ep \gamma)$ kinematics are entirely determined by the five variables $(q_{c.m.}, \epsilon, q'_{c.m.}, \theta'_{c.m.}, \varphi'_{c.m.}$).
}
\label{tab-cs-1}
\begin{tabular} {  rrrr   r   rrrrr rrrrr rrrrr  }
\hline  \hline
\multicolumn{20}{c}{ at fixed \  $q_{c.m.}=1.080$ GeV/c, fixed \ $\epsilon = 0.950$ and fixed  $Q^2=0.92$ GeV$^2$  } \\
\hline 
 $\theta_{c.m.}$ &   $\varphi$ \ \  &  \ \ $\theta'_{c.m.}$ &  \ \ $\varphi'_{c.m.}$  & \ \ &   \multicolumn{5}{c}{$d^5 \sigma \pm \Delta \sigma_{stat}$ } &  \multicolumn{5}{c}{$d^5 \sigma \pm \Delta \sigma_{stat}$ } &  \multicolumn{5}{c}{$d^5 \sigma \pm \Delta \sigma_{stat}$ }   \\ 
  (deg)  &  \ \ (deg)   &   (deg)   &  (deg)  & &    \multicolumn{5}{c}{ \ \ \   at $q'_{c.m.}=45$ MeV/c \ \ \  }  &   \multicolumn{5}{c}{  \ \ \  at $q'_{c.m.}=75$ MeV/c  \ \ \ }  &   \multicolumn{5}{c}{  \ \ \  at $q'_{c.m.}=105$ MeV/c \ \ \  }   \\
\hline 
 50.6 &  82.5 &   40 &      9 &   &  \ \ \ \ &   105.5 &  $\pm$  &    12.0 &  &  \ \ \ \ &    68.0 &  $\pm$  &     5.0 &  &  \ \ \ \ &    48.7 &  $\pm$  &     3.9 &    \\ 
 55.1 &  69.1 &   40 &     27 &   &  \ \ \ \ &   102.6 &  $\pm$  &    17.2 &  &  \ \ \ \ &    75.5 &  $\pm$  &     6.1 &  &  \ \ \ \ &    53.9 &  $\pm$  &     4.5 &    \\ 
 63.0 &  59.3 &   40 &     45 &   &  \ \ \ \ &   141.8 &  $\pm$  &    31.0 &  &  \ \ \ \ &   114.5 &  $\pm$  &    10.2 &  &  \ \ \ \ &    75.0 &  $\pm$  &     6.2 &    \\ 
 73.0 &  53.2 &   40 &     63 &   &  \ \ \ \ &   130.5 &  $\pm$  &    51.3 &  &  \ \ \ \ &   123.9 &  $\pm$  &    16.3 &  &  \ \ \ \ &   117.1 &  $\pm$  &    11.4 &    \\ 
 84.2 &  50.3 &   40 &     81 &   &  \ \ \ \ &   263.7 &  $\pm$  &   116.3 &  &  \ \ \ \ &   238.1 &  $\pm$  &    33.9 &  &  \ \ \ \ &   187.2 &  $\pm$  &    19.3 &    \\ 
 95.8 &  50.3 &   40 &     99 &   &  \ \ \ \ &   536.3 &  $\pm$  &   233.1 &  &  \ \ \ \ &   417.2 &  $\pm$  &    62.5 &  &  \ \ \ \ &   243.8 &  $\pm$  &    28.2 &    \\ 
107.0 &  53.2 &   40 &    117 &   &  \ \ \ \ &  1210.1 &  $\pm$  &   425.1 &  &  \ \ \ \ &   445.1 &  $\pm$  &    67.2 &  &  \ \ \ \ &   300.8 &  $\pm$  &    32.1 &    \\ 
117.0 &  59.3 &   40 &    135 &   &  \ \ \ \ &  1389.4 &  $\pm$  &   306.0 &  &  \ \ \ \ &   491.4 &  $\pm$  &    59.2 &  &  \ \ \ \ &   291.8 &  $\pm$  &    25.2 &    \\ 
124.9 &  69.1 &   40 &    153 &   &  \ \ \ \ &   619.3 &  $\pm$  &   127.0 &  &  \ \ \ \ &   458.1 &  $\pm$  &    39.1 &  &  \ \ \ \ &   247.0 &  $\pm$  &    15.8 &    \\ 
129.4 &  82.5 &   40 &    171 &   &  \ \ \ \ &   570.7 &  $\pm$  &    80.2 &  &  \ \ \ \ &   340.9 &  $\pm$  &    21.7 &  &  \ \ \ \ &   219.7 &  $\pm$  &    10.5 &    \\ 
129.4 &  97.5 &   40 &   -171 &   &  \ \ \ \ &   556.9 &  $\pm$  &    50.4 &  &  \ \ \ \ &   318.1 &  $\pm$  &    13.9 &  &  \ \ \ \ &   204.4 &  $\pm$  &     8.4 &    \\ 
124.9 & 110.9 &   40 &   -153 &   &  \ \ \ \ &   576.2 &  $\pm$  &    34.1 &  &  \ \ \ \ &   286.3 &  $\pm$  &    10.3 &  &  \ \ \ \ &   191.9 &  $\pm$  &     9.2 &    \\ 
117.0 & 120.7 &   40 &   -135 &   &  \ \ \ \ &   443.0 &  $\pm$  &    21.2 &  &  \ \ \ \ &   248.7 &  $\pm$  &     8.9 &  &  \ \ \ \ &   149.2 &  $\pm$  &     8.6 &    \\ 
107.0 & 126.8 &   40 &   -117 &   &  \ \ \ \ &   367.7 &  $\pm$  &    15.9 &  &  \ \ \ \ &   196.3 &  $\pm$  &     8.0 &  &  \ \ \ \ &   130.4 &  $\pm$  &     8.1 &    \\ 
 95.8 & 129.7 &   40 &    -99 &   &  \ \ \ \ &   281.8 &  $\pm$  &    12.7 &  &  \ \ \ \ &   155.0 &  $\pm$  &     7.1 &  &  \ \ \ \ &    93.2 &  $\pm$  &     6.1 &    \\ 
 84.2 & 129.7 &   40 &    -81 &   &  \ \ \ \ &   242.8 &  $\pm$  &    11.7 &  &  \ \ \ \ &   119.6 &  $\pm$  &     6.0 &  &  \ \ \ \ &    88.7 &  $\pm$  &     5.4 &    \\ 
 73.0 & 126.8 &   40 &    -63 &   &  \ \ \ \ &   166.5 &  $\pm$  &     9.6 &  &  \ \ \ \ &   106.4 &  $\pm$  &     5.5 &  &  \ \ \ \ &    57.1 &  $\pm$  &     3.9 &    \\ 
 63.0 & 120.7 &   40 &    -45 &   &  \ \ \ \ &   120.2 &  $\pm$  &     8.3 &  &  \ \ \ \ &    73.8 &  $\pm$  &     4.3 &  &  \ \ \ \ &    50.7 &  $\pm$  &     3.6 &    \\ 
 55.1 & 110.9 &   40 &    -27 &   &  \ \ \ \ &   108.3 &  $\pm$  &     8.6 &  &  \ \ \ \ &    60.9 &  $\pm$  &     4.2 &  &  \ \ \ \ &    48.5 &  $\pm$  &     3.5 &    \\ 
 50.6 &  97.5 &   40 &     -9 &   &  \ \ \ \ &   107.7 &  $\pm$  &     9.5 &  &  \ \ \ \ &    55.8 &  $\pm$  &     4.0 &  &  \ \ \ \ &    36.5 &  $\pm$  &     3.1 &    \\ 
\hline 
  9.0 &   0.0 &   90 &      9 &   &  \ \ \ \ &     9.5 &  $\pm$  &     2.4 &  &  \ \ \ \ &    15.6 &  $\pm$  &     2.3 &  &  \ \ \ \ &    23.1 &  $\pm$  &     3.3 &    \\ 
 27.0 &   0.0 &   90 &     27 &   &  \ \ \ \ &     7.2 &  $\pm$  &     3.0 &  &  \ \ \ \ &    12.8 &  $\pm$  &     2.2 &  &  \ \ \ \ &    12.6 &  $\pm$  &     2.3 &    \\ 
153.0 &   0.0 &   90 &    153 &   &  \ \ \ \ &    33.6 &  $\pm$  &     8.6 &  &  \ \ \ \ &    11.9 &  $\pm$  &     3.5 &  &  \ \ \ \ &     8.8 &  $\pm$  &     2.7 &    \\ 
171.0 &   0.0 &   90 &    171 &   &  \ \ \ \ &    11.3 &  $\pm$  &     2.5 &  &  \ \ \ \ &    10.6 &  $\pm$  &     1.4 &  &  \ \ \ \ &     9.8 &  $\pm$  &     1.0 &    \\ 
171.0 & 180.0 &   90 &   -171 &   &  \ \ \ \ &   258.4 &  $\pm$  &    25.8 &  &  \ \ \ \ &   128.6 &  $\pm$  &     7.6 &  &  \ \ \ \ &    89.6 &  $\pm$  &     5.1 &    \\ 
153.0 & 180.0 &   90 &   -153 &   &  \ \ \ \ &   553.1 &  $\pm$  &    30.7 &  &  \ \ \ \ &   286.1 &  $\pm$  &    10.5 &  &  \ \ \ \ &   178.7 &  $\pm$  &    10.0 &    \\ 
135.0 & 180.0 &   90 &   -135 &   &  \ \ \ \ &   468.1 &  $\pm$  &    18.8 &  &  \ \ \ \ &   266.0 &  $\pm$  &    10.1 &  &  \ \ \ \ &   167.6 &  $\pm$  &    11.7 &    \\ 
117.0 & 180.0 &   90 &   -117 &   &  \ \ \ \ &   402.3 &  $\pm$  &    14.1 &  &  \ \ \ \ &   231.1 &  $\pm$  &     9.8 &  &  \ \ \ \ &   149.4 &  $\pm$  &    10.3 &    \\ 
 99.0 & 180.0 &   90 &    -99 &   &  \ \ \ \ &   286.8 &  $\pm$  &    11.2 &  &  \ \ \ \ &   159.0 &  $\pm$  &     8.1 &  &  \ \ \ \ &    99.3 &  $\pm$  &     6.3 &    \\ 
 81.0 & 180.0 &   90 &    -81 &   &  \ \ \ \ &   194.5 &  $\pm$  &     9.0 &  &  \ \ \ \ &    95.6 &  $\pm$  &     5.4 &  &  \ \ \ \ &    62.6 &  $\pm$  &     4.0 &    \\ 
 63.0 & 180.0 &   90 &    -63 &   &  \ \ \ \ &   121.8 &  $\pm$  &     7.2 &  &  \ \ \ \ &    62.4 &  $\pm$  &     4.1 &  &  \ \ \ \ &    50.8 &  $\pm$  &     3.6 &    \\ 
 45.0 & 180.0 &   90 &    -45 &   &  \ \ \ \ &    74.1 &  $\pm$  &     5.7 &  &  \ \ \ \ &    46.6 &  $\pm$  &     3.3 &  &  \ \ \ \ &    30.2 &  $\pm$  &     2.8 &    \\ 
 27.0 & 180.0 &   90 &    -27 &   &  \ \ \ \ &    39.7 &  $\pm$  &     4.1 &  &  \ \ \ \ &    31.4 &  $\pm$  &     2.6 &  &  \ \ \ \ &    23.2 &  $\pm$  &     2.8 &    \\ 
  9.0 & 180.0 &   90 &     -9 &   &  \ \ \ \ &    17.7 &  $\pm$  &     2.7 &  &  \ \ \ \ &    27.5 &  $\pm$  &     2.8 &  &  \ \ \ \ &    18.6 &  $\pm$  &     2.7 &    \\ 
\hline 
\hline 
\end{tabular}
\end{table*}


\begin{table*}[h]
\caption{Same as previous Table but for data set II below the pion threshold. }
\label{tab-cs-2}
\begin{tabular} {  rrrr   r   rrrrr rrrrr rrrrr  }
\hline  \hline
\multicolumn{20}{c}{ at fixed \  $q_{c.m.}=1.625$ GeV/c , fixed  $\epsilon = 0.879$ and fixed  $Q^2=1.76$  GeV$^2$  } \\
\hline 
 $\theta_{c.m.}$ &   $\varphi$ \ \  &  \ \ $\theta'_{c.m.}$ &  \ \ $\varphi'_{c.m.}$  & \ \ &   \multicolumn{5}{c}{$d^5 \sigma \pm \Delta \sigma_{stat}$ } &  \multicolumn{5}{c}{$d^5 \sigma \pm \Delta \sigma_{stat}$ } &  \multicolumn{5}{c}{$d^5 \sigma \pm \Delta \sigma_{stat}$ }   \\ 
  (deg)  &  \ \ (deg)   &   (deg)   &  (deg)  & &    \multicolumn{5}{c}{ \ \ \   at $q'_{c.m.}=45$ MeV/c \ \ \  }  &   \multicolumn{5}{c}{  \ \ \  at $q'_{c.m.}=75$ MeV/c  \ \ \ }  &   \multicolumn{5}{c}{  \ \ \  at $q'_{c.m.}=105$ MeV/c \ \ \  }   \\
\hline 
 50.6 &  82.5 &   40 &      9 &   &  \ \ \ \ &     9.3 &  $\pm$  &     1.2 &  &  \ \ \ \ &     5.6 &  $\pm$  &     0.6 &  &  \ \ \ \ &     2.9 &  $\pm$  &     0.4 &    \\ 
 55.1 &  69.1 &   40 &     27 &   &  \ \ \ \ &     8.0 &  $\pm$  &     1.5 &  &  \ \ \ \ &     5.0 &  $\pm$  &     0.7 &  &  \ \ \ \ &     3.7 &  $\pm$  &     0.6 &    \\ 
 63.0 &  59.3 &   40 &     45 &   &  \ \ \ \ &     6.9 &  $\pm$  &     2.0 &  &  \ \ \ \ &     7.1 &  $\pm$  &     1.2 &  &  \ \ \ \ &     5.4 &  $\pm$  &     0.9 &    \\ 
 73.0 &  53.2 &   40 &     63 &   &  \ \ \ \ &    16.5 &  $\pm$  &     4.8 &  &  \ \ \ \ &     8.8 &  $\pm$  &     1.7 &  &  \ \ \ \ &     7.6 &  $\pm$  &     1.6 &    \\ 
 84.2 &  50.3 &   40 &     81 &   &  \ \ \ \ &    10.3 &  $\pm$  &     5.6 &  &  \ \ \ \ &    15.7 &  $\pm$  &     3.3 &  &  \ \ \ \ &     9.1 &  $\pm$  &     1.9 &    \\ 
 95.8 &  50.3 &   40 &     99 &   &  \ \ \ \ &    80.2 &  $\pm$  &    21.1 &  &  \ \ \ \ &    30.5 &  $\pm$  &     5.4 &  &  \ \ \ \ &    25.5 &  $\pm$  &     4.1 &    \\ 
107.0 &  53.2 &   40 &    117 &   &  \ \ \ \ &    87.2 &  $\pm$  &    23.2 &  &  \ \ \ \ &    40.1 &  $\pm$  &     6.8 &  &  \ \ \ \ &    26.1 &  $\pm$  &     4.1 &    \\ 
117.0 &  59.3 &   40 &    135 &   &  \ \ \ \ &   122.5 &  $\pm$  &    25.9 &  &  \ \ \ \ &    42.5 &  $\pm$  &     6.0 &  &  \ \ \ \ &    31.1 &  $\pm$  &     3.8 &    \\ 
124.9 &  69.1 &   40 &    153 &   &  \ \ \ \ &   140.5 &  $\pm$  &    25.3 &  &  \ \ \ \ &    51.1 &  $\pm$  &     6.1 &  &  \ \ \ \ &    27.0 &  $\pm$  &     2.8 &    \\ 
129.4 &  82.5 &   40 &    171 &   &  \ \ \ \ &    62.2 &  $\pm$  &    12.0 &  &  \ \ \ \ &    38.9 &  $\pm$  &     3.7 &  &  \ \ \ \ &    25.1 &  $\pm$  &     2.3 &    \\ 
129.4 &  97.5 &   40 &   -171 &   &  \ \ \ \ &    74.3 &  $\pm$  &     9.2 &  &  \ \ \ \ &    34.3 &  $\pm$  &     2.3 &  &  \ \ \ \ &    20.1 &  $\pm$  &     1.5 &    \\ 
124.9 & 110.9 &   40 &   -153 &   &  \ \ \ \ &    51.7 &  $\pm$  &     4.6 &  &  \ \ \ \ &    33.1 &  $\pm$  &     1.6 &  &  \ \ \ \ &    18.6 &  $\pm$  &     0.9 &    \\ 
117.0 & 120.7 &   40 &   -135 &   &  \ \ \ \ &    42.4 &  $\pm$  &     2.9 &  &  \ \ \ \ &    25.4 &  $\pm$  &     1.2 &  &  \ \ \ \ &    15.3 &  $\pm$  &     0.8 &    \\ 
107.0 & 126.8 &   40 &   -117 &   &  \ \ \ \ &    36.1 &  $\pm$  &     2.2 &  &  \ \ \ \ &    18.9 &  $\pm$  &     0.9 &  &  \ \ \ \ &    10.6 &  $\pm$  &     0.7 &    \\ 
 95.8 & 129.7 &   40 &    -99 &   &  \ \ \ \ &    28.4 &  $\pm$  &     1.7 &  &  \ \ \ \ &    12.8 &  $\pm$  &     0.7 &  &  \ \ \ \ &     7.0 &  $\pm$  &     0.6 &    \\ 
 84.2 & 129.7 &   40 &    -81 &   &  \ \ \ \ &    19.4 &  $\pm$  &     1.6 &  &  \ \ \ \ &    10.1 &  $\pm$  &     0.7 &  &  \ \ \ \ &     6.5 &  $\pm$  &     0.6 &    \\ 
 73.0 & 126.8 &   40 &    -63 &   &  \ \ \ \ &    14.5 &  $\pm$  &     1.5 &  &  \ \ \ \ &     7.2 &  $\pm$  &     0.6 &  &  \ \ \ \ &     3.7 &  $\pm$  &     0.4 &    \\ 
 63.0 & 120.7 &   40 &    -45 &   &  \ \ \ \ &    10.6 &  $\pm$  &     1.2 &  &  \ \ \ \ &     5.5 &  $\pm$  &     0.5 &  &  \ \ \ \ &     3.9 &  $\pm$  &     0.5 &    \\ 
 55.1 & 110.9 &   40 &    -27 &   &  \ \ \ \ &     8.4 &  $\pm$  &     1.1 &  &  \ \ \ \ &     6.3 &  $\pm$  &     0.6 &  &  \ \ \ \ &     3.2 &  $\pm$  &     0.5 &    \\ 
 50.6 &  97.5 &   40 &     -9 &   &  \ \ \ \ &     9.0 &  $\pm$  &     1.3 &  &  \ \ \ \ &     3.9 &  $\pm$  &     0.5 &  &  \ \ \ \ &     3.6 &  $\pm$  &     0.5 &    \\ 
\hline 
  9.0 &   0.0 &   90 &      9 &   &  \ \ \ \ &     1.5 &  $\pm$  &     0.5 &  &  \ \ \ \ &     1.3 &  $\pm$  &     0.4 &  &  \ \ \ \ &     1.1 &  $\pm$  &     0.3 &    \\ 
 27.0 &   0.0 &   90 &     27 &   &  \ \ \ \ &     1.2 &  $\pm$  &     0.7 &  &  \ \ \ \ &     0.7 &  $\pm$  &     0.3 &  &  \ \ \ \ &     1.3 &  $\pm$  &     0.4 &    \\ 
171.0 &   0.0 &   90 &    171 &   &  \ \ \ \ &     5.3 &  $\pm$  &     1.2 &  &  \ \ \ \ &     1.6 &  $\pm$  &     0.4 &  &  \ \ \ \ &     0.9 &  $\pm$  &     0.2 &    \\ 
171.0 & 180.0 &   90 &   -171 &   &  \ \ \ \ &    39.9 &  $\pm$  &     7.3 &  &  \ \ \ \ &    25.5 &  $\pm$  &     2.4 &  &  \ \ \ \ &    15.8 &  $\pm$  &     1.1 &    \\ 
153.0 & 180.0 &   90 &   -153 &   &  \ \ \ \ &    64.1 &  $\pm$  &     6.0 &  &  \ \ \ \ &    44.7 &  $\pm$  &     2.1 &  &  \ \ \ \ &    26.4 &  $\pm$  &     1.1 &    \\ 
135.0 & 180.0 &   90 &   -135 &   &  \ \ \ \ &    65.6 &  $\pm$  &     3.7 &  &  \ \ \ \ &    32.2 &  $\pm$  &     1.3 &  &  \ \ \ \ &    19.9 &  $\pm$  &     0.9 &    \\ 
117.0 & 180.0 &   90 &   -117 &   &  \ \ \ \ &    42.6 &  $\pm$  &     2.1 &  &  \ \ \ \ &    20.8 &  $\pm$  &     0.9 &  &  \ \ \ \ &    12.7 &  $\pm$  &     0.8 &    \\ 
 99.0 & 180.0 &   90 &    -99 &   &  \ \ \ \ &    24.4 &  $\pm$  &     1.3 &  &  \ \ \ \ &    12.8 &  $\pm$  &     0.7 &  &  \ \ \ \ &     8.3 &  $\pm$  &     0.7 &    \\ 
 81.0 & 180.0 &   90 &    -81 &   &  \ \ \ \ &    14.1 &  $\pm$  &     0.9 &  &  \ \ \ \ &     8.2 &  $\pm$  &     0.6 &  &  \ \ \ \ &     4.5 &  $\pm$  &     0.5 &    \\ 
 63.0 & 180.0 &   90 &    -63 &   &  \ \ \ \ &    10.7 &  $\pm$  &     0.8 &  &  \ \ \ \ &     5.5 &  $\pm$  &     0.5 &  &  \ \ \ \ &     3.5 &  $\pm$  &     0.5 &    \\ 
 45.0 & 180.0 &   90 &    -45 &   &  \ \ \ \ &     7.4 &  $\pm$  &     0.8 &  &  \ \ \ \ &     3.8 &  $\pm$  &     0.4 &  &  \ \ \ \ &     2.1 &  $\pm$  &     0.4 &    \\ 
 27.0 & 180.0 &   90 &    -27 &   &  \ \ \ \ &     3.4 &  $\pm$  &     0.6 &  &  \ \ \ \ &     3.2 &  $\pm$  &     0.4 &  &  \ \ \ \ &     2.1 &  $\pm$  &     0.4 &    \\ 
  9.0 & 180.0 &   90 &     -9 &   &  \ \ \ \ &     2.8 &  $\pm$  &     0.7 &  &  \ \ \ \ &     1.8 &  $\pm$  &     0.4 &  &  \ \ \ \ &     0.9 &  $\pm$  &     0.3 &    \\ 
\hline 
\hline 
\end{tabular}
\end{table*}


\begin{table*}[h]
\caption{Same as previous Table but for data set I-a  above the pion threshold. }
\label{tab-cs-3}
\begin{tabular} {  rrrr   r   rrrrr rrrrr rrrrr  }
\hline  \hline
\multicolumn{20}{c}{ at fixed \  $q_{c.m.}=1.080$ GeV/c , fixed  $\epsilon = 0.950$ and fixed  $Q^2=0.92$  GeV$^2$  } \\
\hline 
 $\theta_{c.m.}$ &   $\varphi$ \ \  &  \ \ $\theta'_{c.m.}$ &  \ \ $\varphi'_{c.m.}$  & \ \ &   \multicolumn{5}{c}{$d^5 \sigma \pm \Delta \sigma_{stat}$ } &  \multicolumn{5}{c}{$d^5 \sigma \pm \Delta \sigma_{stat}$ } &  \multicolumn{5}{c}{$d^5 \sigma \pm \Delta \sigma_{stat}$ }   \\ 
  (deg)  &  \ \ (deg)   &   (deg)   &  (deg)  &  &   \multicolumn{5}{c}{  \ \ \   at $q'_{c.m.}=135$ MeV/c \ \ \  }  &   \multicolumn{5}{c}{  \ \ \  at $q'_{c.m.}=165$ MeV/c \ \ \  }  &   \multicolumn{5}{c}{  \ \ \  at $q'_{c.m.}=215$ MeV/c \ \ \  }   \\
\hline 
 50.6 &  82.5 &   40 &      9 &  &   &    38.8 &  $\pm$  &     5.2 &  &   &    32.8 &  $\pm$  &    20.6 &  &   & &  &   \\ 
 55.1 &  69.1 &   40 &     27 &  &   &    44.5 &  $\pm$  &     5.9 &  &   &    53.6 &  $\pm$  &    18.1 &  &   & &  &   \\ 
 63.0 &  59.3 &   40 &     45 &  &   &    54.2 &  $\pm$  &     6.1 &  &   &    43.8 &  $\pm$  &    10.1 &  &   & &  &   \\ 
 73.0 &  53.2 &   40 &     63 &  &   &    91.4 &  $\pm$  &    11.6 &  &   &    97.5 &  $\pm$  &    21.9 &  &   & &  &   \\ 
 84.2 &  50.3 &   40 &     81 &  &   &    91.1 &  $\pm$  &    15.5 &  &   &   168.3 &  $\pm$  &    42.7 &  &   & &  &   \\ 
 95.8 &  50.3 &   40 &     99 &   &  \ \ \ \ &   175.5 &  $\pm$  &    26.2 &  &  \ \ \ \ &   143.0 &  $\pm$  &    38.8 &  &  \ \ \ \ &   159.6 &  $\pm$  &    95.1 &    \\ 
107.0 &  53.2 &   40 &    117 &   &  \ \ \ \ &   229.0 &  $\pm$  &    28.7 &  &  \ \ \ \ &   169.4 &  $\pm$  &    33.4 &  &  \ \ \ \ &   168.8 &  $\pm$  &    72.4 &    \\ 
117.0 &  59.3 &   40 &    135 &   &  \ \ \ \ &   193.8 &  $\pm$  &    19.9 &  &  \ \ \ \ &   153.2 &  $\pm$  &    19.6 &  &  \ \ \ \ &   157.1 &  $\pm$  &    24.9 &    \\ 
124.9 &  69.1 &   40 &    153 &   &  \ \ \ \ &   184.4 &  $\pm$  &    14.0 &  &  \ \ \ \ &   124.1 &  $\pm$  &    11.3 &  &  \ \ \ \ &   119.7 &  $\pm$  &    14.6 &    \\ 
129.4 &  82.5 &   40 &    171 &   &  \ \ \ \ &   178.5 &  $\pm$  &    12.1 &  &  \ \ \ \ &   125.2 &  $\pm$  &    12.6 &  &  \ \ \ \ &   125.5 &  $\pm$  &    18.3 &    \\ 
129.4 &  97.5 &   40 &   -171 &   &  \ \ \ \ &   121.4 &  $\pm$  &     9.9 &  &  \ \ \ \ &   127.1 &  $\pm$  &    14.3 &  &  \ \ \ \ &   155.6 &  $\pm$  &    39.6 &    \\ 
124.9 & 110.9 &   40 &   -153 &  &   &   153.8 &  $\pm$  &    13.4 &  &   &    81.2 &  $\pm$  &    19.4 &  &   & &  &   \\ 
117.0 & 120.7 &   40 &   -135 &  &   &    90.2 &  $\pm$  &    11.3 &  &   &   141.0 &  $\pm$  &    29.3 &  &   & &  &   \\ 
107.0 & 126.8 &   40 &   -117 &  &   &    74.7 &  $\pm$  &     8.6 &  &   &    80.7 &  $\pm$  &    28.3 &  &   & &  &   \\ 
 95.8 & 129.7 &   40 &    -99 &  &   &    67.3 &  $\pm$  &     7.0 &  &   &    71.9 &  $\pm$  &    33.1 &  &   & &  &   \\ 
 84.2 & 129.7 &   40 &    -81 &  &   &    69.9 &  $\pm$  &     8.2 &  &   &    36.0 &  $\pm$  &    33.9 &  &   & &  &   \\ 
 73.0 & 126.8 &   40 &    -63 &  &  &    32.2 &  $\pm$ &      4.7 &  &  &  & & & & & & & &       \\ 
 63.0 & 120.7 &   40 &    -45 &  &  &    45.5 &  $\pm$ &      6.0 &  &  &  & & & & & & & &       \\ 
 55.1 & 110.9 &   40 &    -27 &  &  &    33.9 &  $\pm$ &      5.3 &  &  &  & & & & & & & &       \\ 
 50.6 &  97.5 &   40 &     -9 &  &  &    41.2 &  $\pm$ &      5.9 &  &  &  & & & & & & & &       \\ 
\hline 
  9.0 &   0.0 &   90 &      9 &  &  &    14.9 &  $\pm$ &      5.6 &  &  &  & & & & & & & &       \\ 
 27.0 &   0.0 &   90 &     27 &  &  &    22.7 &  $\pm$ &      6.2 &  &  &  & & & & & & & &       \\ 
 45.0 &   0.0 &   90 &     45 &  &  &    18.1 &  $\pm$ &      8.2 &  &  &  & & & & & & & &       \\ 
135.0 &   0.0 &   90 &    135 &  &  &  & & & & & & & & &  &   176.9  & $\pm$  &    75.4 &  \\ 
153.0 &   0.0 &   90 &    153 &   &  \ \ \ \ &     7.2 &  $\pm$  &     1.7 &  &  \ \ \ \ &    16.5 &  $\pm$  &     3.0 &  &  \ \ \ \ &    80.1 &  $\pm$  &    12.6 &    \\ 
171.0 &   0.0 &   90 &    171 &   &  \ \ \ \ &    14.6 &  $\pm$  &     1.8 &  &  \ \ \ \ &    31.1 &  $\pm$  &     4.5 &  &  \ \ \ \ &    79.7 &  $\pm$  &    17.9 &    \\ 
171.0 & 180.0 &   90 &   -171 &   &  \ \ \ \ &    84.5 &  $\pm$  &     7.8 &  &  \ \ \ \ &    94.0 &  $\pm$  &    11.8 &  &  \ \ \ \ &   216.1 &  $\pm$  &    70.3 &    \\ 
153.0 & 180.0 &   90 &   -153 &  &   &   137.6 &  $\pm$  &    12.8 &  &   &   176.5 &  $\pm$  &    52.7 &  &   & &  &   \\ 
135.0 & 180.0 &   90 &   -135 &  &  &   112.2 &  $\pm$ &     16.3 &  &  &  & & & & & & & &       \\ 
117.0 & 180.0 &   90 &   -117 &  &  &    61.8 &  $\pm$ &      9.8 &  &  &  & & & & & & & &       \\ 
 99.0 & 180.0 &   90 &    -99 &  &  &    56.3 &  $\pm$ &     14.4 &  &  &  & & & & & & & &       \\ 
 81.0 & 180.0 &   90 &    -81 &  &  &    45.8 &  $\pm$ &     20.3 &  &  &  & & & & & & & &       \\ 
 27.0 & 180.0 &   90 &    -27 &  &  &    28.1 &  $\pm$ &     10.4 &  &  &  & & & & & & & &       \\ 
  9.0 & 180.0 &   90 &     -9 &  &  &    30.8 &  $\pm$ &      9.3 &  &  &  & & & & & & & &       \\ 
\hline 
\hline 
\end{tabular}
\end{table*}


\begin{table*}[h]
\caption{Same as previous Table but for data set II  above the pion threshold.}
\label{tab-cs-4}
\begin{tabular} {  rrrr   r   rrrrr rrrrr rrrrr  }
\hline  \hline
\multicolumn{20}{c}{ at fixed \  $q_{c.m.}=1.625$ GeV/c , fixed  $\epsilon = 0.879$ and fixed  $Q^2=1.76$  GeV$^2$  } \\
\hline 
 $\theta_{c.m.}$ &   $\varphi$ \ \  &  \ \ $\theta'_{c.m.}$ &  \ \ $\varphi'_{c.m.}$  & \ \ &   \multicolumn{5}{c}{$d^5 \sigma \pm \Delta \sigma_{stat}$ } &  \multicolumn{5}{c}{$d^5 \sigma \pm \Delta \sigma_{stat}$ } &  \multicolumn{5}{c}{$d^5 \sigma \pm \Delta \sigma_{stat}$ }   \\ 
  (deg)  &  \ \ (deg)   &   (deg)   &  (deg)  &  &   \multicolumn{5}{c}{  \ \ \   at $q'_{c.m.}=135$ MeV/c \ \ \  }  &   \multicolumn{5}{c}{  \ \ \  at $q'_{c.m.}=165$ MeV/c \ \ \  }  &   \multicolumn{5}{c}{  \ \ \  at $q'_{c.m.}=215$ MeV/c \ \ \  }   \\
\hline 
 50.6 &  82.5 &   40 &      9 &  &  &     2.3 &  $\pm$ &      0.5 &  &  &  & & & & & & & &       \\ 
 55.1 &  69.1 &   40 &     27 &  &   &     3.8 &  $\pm$  &     0.7 &  &   &     2.8 &  $\pm$  &     1.1 &  &   & &  &   \\ 
 63.0 &  59.3 &   40 &     45 &  &   &     4.1 &  $\pm$  &     0.8 &  &   &     3.4 &  $\pm$  &     1.1 &  &   & &  &   \\ 
 73.0 &  53.2 &   40 &     63 &  &   &     5.0 &  $\pm$  &     1.4 &  &   &     5.3 &  $\pm$  &     2.0 &  &   & &  &   \\ 
 84.2 &  50.3 &   40 &     81 &  &   &     7.9 &  $\pm$  &     2.1 &  &   &     4.2 &  $\pm$  &     2.4 &  &   & &  &   \\ 
 95.8 &  50.3 &   40 &     99 &   &  \ \ \ \ &    10.8 &  $\pm$  &     2.6 &  &  \ \ \ \ &    19.7 &  $\pm$  &     5.9 &  &  \ \ \ \ &     9.3 &  $\pm$  &     5.1 &    \\ 
107.0 &  53.2 &   40 &    117 &   &  \ \ \ \ &    18.3 &  $\pm$  &     2.8 &  &  \ \ \ \ &     9.1 &  $\pm$  &     2.1 &  &  \ \ \ \ &    14.4 &  $\pm$  &     3.4 &    \\ 
117.0 &  59.3 &   40 &    135 &   &  \ \ \ \ &    21.3 &  $\pm$  &     2.7 &  &  \ \ \ \ &    13.9 &  $\pm$  &     2.3 &  &  \ \ \ \ &     9.9 &  $\pm$  &     2.1 &    \\ 
124.9 &  69.1 &   40 &    153 &   &  \ \ \ \ &    18.0 &  $\pm$  &     2.0 &  &  \ \ \ \ &    15.0 &  $\pm$  &     1.8 &  &  \ \ \ \ &    12.3 &  $\pm$  &     1.6 &    \\ 
129.4 &  82.5 &   40 &    171 &   &  \ \ \ \ &    17.1 &  $\pm$  &     1.3 &  &  \ \ \ \ &    13.6 &  $\pm$  &     1.1 &  &  \ \ \ \ &     8.8 &  $\pm$  &     0.8 &    \\ 
129.4 &  97.5 &   40 &   -171 &   &  \ \ \ \ &    18.2 &  $\pm$  &     1.0 &  &  \ \ \ \ &    13.2 &  $\pm$  &     0.9 &  &  \ \ \ \ &     8.6 &  $\pm$  &     0.7 &    \\ 
124.9 & 110.9 &   40 &   -153 &   &  \ \ \ \ &    14.3 &  $\pm$  &     0.8 &  &  \ \ \ \ &     9.7 &  $\pm$  &     0.8 &  &  \ \ \ \ &     8.0 &  $\pm$  &     1.2 &    \\ 
117.0 & 120.7 &   40 &   -135 &   &  \ \ \ \ &     9.7 &  $\pm$  &     0.7 &  &  \ \ \ \ &     9.5 &  $\pm$  &     1.1 &  &  \ \ \ \ &     8.7 &  $\pm$  &     2.6 &    \\ 
107.0 & 126.8 &   40 &   -117 &  &   &     7.5 &  $\pm$  &     0.7 &  &   &     5.0 &  $\pm$  &     1.0 &  &   & &  &   \\ 
 95.8 & 129.7 &   40 &    -99 &  &   &     6.2 &  $\pm$  &     0.7 &  &   &     8.0 &  $\pm$  &     2.2 &  &   & &  &   \\ 
 84.2 & 129.7 &   40 &    -81 &  &   &     5.6 &  $\pm$  &     0.8 &  &   &     2.6 &  $\pm$  &     1.2 &  &   & &  &   \\ 
 73.0 & 126.8 &   40 &    -63 &  &   &     4.0 &  $\pm$  &     0.7 &  &   &     4.9 &  $\pm$  &     2.9 &  &   & &  &   \\ 
 63.0 & 120.7 &   40 &    -45 &  &   &     3.3 &  $\pm$  &     0.7 &  &   &     5.2 &  $\pm$  &     3.0 &  &   & &  &   \\ 
 55.1 & 110.9 &   40 &    -27 &  &   &     2.0 &  $\pm$  &     0.5 &  &   &     2.3 &  $\pm$  &     1.5 &  &   & &  &   \\ 
 50.6 &  97.5 &   40 &     -9 &  &  &     3.9 &  $\pm$ &      0.9 &  &  &  & & & & & & & &       \\ 
\hline 
  9.0 &   0.0 &   90 &      9 &  &  &     1.9 &  $\pm$ &      1.0 &  &  &  & & & & & & & &       \\ 
 27.0 &   0.0 &   90 &     27 &  &  &     0.7 &  $\pm$ &      0.4 &  &  &  & & & & & & & &       \\ 
153.0 &   0.0 &   90 &    153 &  &  &  & & & & & & & & &  &     3.7  & $\pm$  &     1.3 &  \\ 
171.0 &   0.0 &   90 &    171 &   &  \ \ \ \ &     1.4 &  $\pm$  &     0.2 &  &  \ \ \ \ &     1.7 &  $\pm$  &     0.2 &  &  \ \ \ \ &     4.2 &  $\pm$  &     0.4 &    \\ 
171.0 & 180.0 &   90 &   -171 &   &  \ \ \ \ &    12.7 &  $\pm$  &     0.8 &  &  \ \ \ \ &    10.1 &  $\pm$  &     0.7 &  &  \ \ \ \ &     9.5 &  $\pm$  &     0.7 &    \\ 
153.0 & 180.0 &   90 &   -153 &   &  \ \ \ \ &    19.1 &  $\pm$  &     0.9 &  &  \ \ \ \ &    14.4 &  $\pm$  &     1.0 &  &  \ \ \ \ &    13.6 &  $\pm$  &     1.9 &    \\ 
135.0 & 180.0 &   90 &   -135 &  &   &    13.1 &  $\pm$  &     0.8 &  &   &    10.3 &  $\pm$  &     1.5 &  &   & &  &   \\ 
117.0 & 180.0 &   90 &   -117 &  &  &     8.6 &  $\pm$ &      0.9 &  &  &  & & & & & & & &       \\ 
 99.0 & 180.0 &   90 &    -99 &  &  &     6.2 &  $\pm$ &      1.1 &  &  &  & & & & & & & &       \\ 
 81.0 & 180.0 &   90 &    -81 &  &  &     1.3 &  $\pm$ &      0.7 &  &  &  & & & & & & & &       \\ 
 63.0 & 180.0 &   90 &    -63 &  &  &     4.3 &  $\pm$ &      2.6 &  &  &  & & & & & & & &       \\ 
  9.0 & 180.0 &   90 &     -9 &  &  &     3.5 &  $\pm$ &      1.7 &  &  &  & & & & & & & &       \\ 
\hline 
\hline 
\end{tabular}
\end{table*}


\begin{table*}
\caption{ 
The measured $(ep \to ep \gamma)$ cross section $d^5 \sigma / d k'_{e lab} d \Omega'_{e lab} d\Omega_{\gamma c.m.}$ ($\pm$ statistical error $\pm$ systematic error) in pb/(GeV sr$^2$), for data set I-b.  The  $(ep \to ep \gamma)$ kinematics are entirely determined by the five variables:  $Q^2$ (= 1.0 GeV$^2$ fixed), beam energy $E_{beam}$ (= 4.032 GeV fixed), and ($W, \cos \theta_{c.m.}$, $\varphi$) which are given in the Table. Numbers in parenthesis represent a rough estimate of the systematic error, when missing.
}
\label{tab-cs-5}
\begin{tabular}{ccccccc} \hline \hline
$W$ (GeV) 
& $\varphi=15^\circ$
& $\varphi=45^\circ$
& $\varphi=75^\circ$
& $\varphi=105^\circ$
& $\varphi=135^\circ$
& $\varphi=165^\circ$ \\ 
\hline 
\multicolumn{7}{c}{  $\cos \theta_{c.m.}= -0.975$ } \\
\hline
  1.05 &   \  &   \  &   \  &   \  &   \  &  140 $\pm$   73 $\pm$   62 \\
  1.07 &   \  &   \  &   \  &   69 $\pm$   41 $\pm$   18 &   63 $\pm$   24 $\pm$   25 &  120 $\pm$   30 $\pm$   24 \\
  1.09 &   \  &   \  &   50 $\pm$   32 $\pm$    9 &   91 $\pm$   24 $\pm$   10 &   92 $\pm$   20 $\pm$   12 &   86 $\pm$   19 $\pm$   24 \\
  1.11 &   \  &   94 $\pm$   40 $\pm$    2 &   53 $\pm$   17 $\pm$    9 &  114 $\pm$   20 $\pm$   11 &  110 $\pm$   21 $\pm$    9 &   61 $\pm$   16 $\pm$   18 \\
  1.13 &   \  &   36 $\pm$   14 $\pm$    7 &   54 $\pm$   14 $\pm$   15 &   75 $\pm$   16 $\pm$    8 &   72 $\pm$   17 $\pm$   17 &   53 $\pm$   16 $\pm$    3 \\
  1.15 &   43 $\pm$   14 $\pm$   15 &   95 $\pm$   18 $\pm$   15 &   96 $\pm$   17 $\pm$    9 &   97 $\pm$   18 $\pm$   14 &  131 $\pm$   23 $\pm$   43 &  104 $\pm$   19 $\pm$   27 \\
  1.17 &   55 $\pm$   13 $\pm$    9 &  112 $\pm$   18 $\pm$    8 &  102 $\pm$   17 $\pm$   12 &  124 $\pm$   18 $\pm$    9 &  179 $\pm$   22 $\pm$   24 &  168 $\pm$   22 $\pm$   10 \\
  1.19 &   93 $\pm$   16 $\pm$    7 &  116 $\pm$   17 $\pm$   13 &  136 $\pm$   18 $\pm$   13 &  154 $\pm$   18 $\pm$   23 &  145 $\pm$   19 $\pm$   14 &  178 $\pm$   24 $\pm$   15 \\
  1.21 &  118 $\pm$   18 $\pm$   17 &  116 $\pm$   17 $\pm$   20 &  167 $\pm$   18 $\pm$    9 &  119 $\pm$   15 $\pm$   10 &  152 $\pm$   21 $\pm$    6 &  144 $\pm$   24 $\pm$   20 \\
  1.23 &  111 $\pm$   16 $\pm$   13 &  109 $\pm$   15 $\pm$   18 &  102 $\pm$   13 $\pm$   11 &  141 $\pm$   16 $\pm$    8 &  107 $\pm$   17 $\pm$   18 &   83 $\pm$   15 $\pm$   14 \\
  1.25 &   51 $\pm$   11 $\pm$   12 &   78 $\pm$   12 $\pm$   11 &   94 $\pm$   12 $\pm$   11 &   74 $\pm$   13 $\pm$   10 &   81 $\pm$   11 $\pm$    6 &   96 $\pm$   12 $\pm$   10 \\
  1.27 &   41 $\pm$    9 $\pm$    7 &   51 $\pm$    9 $\pm$    6 &   48 $\pm$    9 $\pm$    6 &   64 $\pm$    9 $\pm$    4 &   61 $\pm$    8 $\pm$    4 &   47 $\pm$    7 $\pm$    5 \\
\hline
\multicolumn{7}{c}{  $\cos \theta_{c.m.}= -0.875$ } \\
\hline
  1.05 &   \  &   \  &   \  &   \  &  296 $\pm$   90 $\pm$   18 &  103 $\pm$   37 $\pm$   11 \\
  1.07 &   \  &   \  &   \  &  169 $\pm$   53 $\pm$   12 &  163 $\pm$   36 $\pm$    6 &  139 $\pm$   30 $\pm$   14 \\
  1.09 &   \  &   \  &  374 $\pm$  163 $\pm$    9 &   98 $\pm$   25 $\pm$   40 &  113 $\pm$   25 $\pm$   14 &  122 $\pm$   29 $\pm$   11 \\
  1.11 &   \  &   \  &  107 $\pm$   32 $\pm$    8 &  129 $\pm$   24 $\pm$   10 &  119 $\pm$   28 $\pm$   13 &   58 $\pm$   23 $\pm$   45 \\
  1.13 &   \  &   \  &   91 $\pm$   21 $\pm$    6 &  123 $\pm$   24 $\pm$    4 &   68 $\pm$   21 $\pm$   16 &   99 $\pm$   31 $\pm$   11 \\
  1.15 &   \  &   81 $\pm$   29 $\pm$   12 &  143 $\pm$   23 $\pm$    9 &   97 $\pm$   20 $\pm$   18 &   99 $\pm$   22 $\pm$   13 &  117 $\pm$   25 $\pm$   10 \\
  1.17 &   \  &   38 $\pm$   15 $\pm$   15 &   99 $\pm$   18 $\pm$    5 &  106 $\pm$   18 $\pm$   11 &  118 $\pm$   22 $\pm$   14 &  106 $\pm$   30 $\pm$   14 \\
  1.19 &  172 $\pm$   44 $\pm$   26 &  112 $\pm$   19 $\pm$    8 &  127 $\pm$   18 $\pm$   11 &  125 $\pm$   18 $\pm$   12 &  145 $\pm$   29 $\pm$   18 &  272 $\pm$  122 $\pm$   90 \\
  1.21 &  103 $\pm$   22 $\pm$   11 &  123 $\pm$   19 $\pm$    7 &  134 $\pm$   17 $\pm$    5 &  183 $\pm$   21 $\pm$   11 &  178 $\pm$   51 $\pm$   18 &  217 $\pm$   70 $\pm$   46 \\
  1.23 &  129 $\pm$   21 $\pm$    8 &  117 $\pm$   18 $\pm$    6 &  135 $\pm$   16 $\pm$    9 &  133 $\pm$   20 $\pm$   12 &  130 $\pm$   24 $\pm$   42 &   68 $\pm$   19 $\pm$   28 \\
  1.25 &   99 $\pm$   17 $\pm$    6 &   95 $\pm$   17 $\pm$    6 &   99 $\pm$   13 $\pm$    7 &   68 $\pm$   14 $\pm$    8 &   88 $\pm$   14 $\pm$   14 &   81 $\pm$   21 $\pm$    8 \\
  1.27 &   54 $\pm$   13 $\pm$    7 &   60 $\pm$   13 $\pm$   14 &   87 $\pm$   13 $\pm$    5 &   71 $\pm$   11 $\pm$    6 &   57 $\pm$   12 $\pm$    7 &   \  \\
\hline
\multicolumn{7}{c}{  $\cos \theta_{c.m.}= -0.650$ } \\
 \hline
  1.03 &   \  &   \  &   \  &   \  &   \  &  287 $\pm$  103 $\pm$   11 \\
  1.05 &   \  &   \  &   \  &  201 $\pm$  177 $\pm$   16 &  211 $\pm$   58 $\pm$   14 &  188 $\pm$   46 $\pm$    9 \\
  1.07 &   \  &   \  &   \  &  191 $\pm$   52 $\pm$   10 &  112 $\pm$   28 $\pm$   11 &  131 $\pm$   31 $\pm$   20 \\
  1.09 &   \  &   \  &   \  &  150 $\pm$   31 $\pm$    9 &  110 $\pm$   26 $\pm$   10 &   72 $\pm$   32 $\pm$  101 \\
  1.11 &   \  &   \  &  114 $\pm$   36 $\pm$    5 &  108 $\pm$   22 $\pm$    7 &  179 $\pm$   46 $\pm$   41 &  153 $\pm$   77 $\pm$ (120) \\
  1.13 &   \  &   \  &  115 $\pm$   25 $\pm$    5 &  139 $\pm$   27 $\pm$    7 &   62 $\pm$   24 $\pm$  515 &   78 $\pm$   32 $\pm$ (120) \\
  1.15 &   \  &   \  &   86 $\pm$   18 $\pm$    5 &  116 $\pm$   23 $\pm$   11 &  128 $\pm$   27 $\pm$  169 &   91 $\pm$   36 $\pm$ (120) \\
  1.17 &   \  &  145 $\pm$   48 $\pm$   13 &  149 $\pm$   21 $\pm$   10 &  137 $\pm$   22 $\pm$   15 &  132 $\pm$   33 $\pm$  400 &   \  \\
  1.19 &   \  &  129 $\pm$   30 $\pm$   24 &  169 $\pm$   21 $\pm$    5 &  152 $\pm$   22 $\pm$   27 &  178 $\pm$   69 $\pm$ (200) &   \  \\
  1.21 &   \  &  175 $\pm$   28 $\pm$   22 &  152 $\pm$   19 $\pm$   14 &  125 $\pm$   21 $\pm$   17 &   \  &   \  \\
  1.23 &   66 $\pm$   33 $\pm$    (30) &  104 $\pm$   19 $\pm$   16 &  131 $\pm$   17 $\pm$   10 &   87 $\pm$   21 $\pm$   10 &   88 $\pm$   38 $\pm$ (200) &   \  \\
  1.25 &   79 $\pm$   23 $\pm$    (20) &  108 $\pm$   19 $\pm$   12 &   76 $\pm$   12 $\pm$   12 &   50 $\pm$   15 $\pm$   59 &  125 $\pm$   30 $\pm$ (200) &   \  \\
  1.27 &   80 $\pm$   19 $\pm$    (20) &  107 $\pm$   20 $\pm$   11 &  102 $\pm$   15 $\pm$   15 &   85 $\pm$   16 $\pm$   73 &   89 $\pm$   42 $\pm$ (200) &   \  \\
\hline \hline
\end{tabular}\end{table*}

\end{document}